\documentclass[usenatbib,fleqn]{mnras}
\usepackage{graphicx,hyperref,amsmath,amssymb,mathrsfs,xspace}
\usepackage[dvipsnames]{xcolor}

\title{Gaia EDR3 view on Galactic globular clusters}

\author[Vasiliev \& Baumgardt]{
Eugene Vasiliev$^{1,2}$\thanks{E-mail: eugvas@lpi.ru}, Holger Baumgardt$^3$\\
$^1$Institute of Astronomy, Madingley road, Cambridge, CB3 0HA, UK\\
$^2$Lebedev Physical Institute, Leninsky prospekt 53, Moscow, 119991, Russia\\
$^3$School of Mathematics and Physics, The University of Queensland, St.Lucia, QLD 4072, Australia
}

\newcommand{\Gaia}{\textit{Gaia}\xspace}
\newcommand{\kms}{km\:s$^{-1}$\xspace}
\newcommand{\masyr}{mas\:yr$^{-1}$\xspace}

\date{Accepted 2021 May 17. Received 2021 May 11; in original form 2021 March 17}
\pagerange{5978--6002}\volume{505}\pubyear{2021}
\begin{document}
\maketitle

\begin{abstract}
We use the data from \Gaia Early Data Release 3 (EDR3) to study the kinematic properties of Milky Way globular clusters. We measure the mean parallaxes and proper motions (PM) for 170 clusters, determine the PM dispersion profiles for more than 100 clusters, uncover rotation signatures in more than 20 objects, and find evidence for radial or tangential PM anisotropy in a dozen richest clusters. At the same time, we use the selection of cluster members to explore the reliability and limitations of the \Gaia catalogue itself. We find that the formal uncertainties on parallax and PM are underestimated by $10-20\%$ in dense central regions even for stars that pass numerous quality filters. We explore the spatial covariance function of systematic errors, and determine a lower limit on the uncertainty of average parallaxes and PM at the level 0.01~mas and 0.025~\masyr, respectively. Finally, a comparison of mean parallaxes of clusters with distances from various literature sources suggests that the parallaxes for stars with $G>13$ (after applying the zero-point correction suggested by \citealt{Lindegren2021b}) are overestimated by $\sim 0.01\pm0.003$~mas. Despite these caveats, the quality of \Gaia astrometry has been significantly improved in EDR3 and provides valuable insights into the properties of star clusters.
\end{abstract}


\section{Introduction}   \label{sec:intro}

The most recent data release (EDR3) from the \Gaia mission \citep{Brown2021} does not provide new data products, but instead improves upon the previous DR2 in various aspects related to the photometric and astrometric catalogues. In particular, the statistical uncertainties on parallaxes $\varpi$ and proper motions (PM) $\boldsymbol\mu$ have been reduced by a factor of two on average, and the systematic uncertainties are reduced even further. Already after DR2, it became possible to measure the mean parallaxes \citep{Chen2018,Shao2019} and PM of almost all Milky Way globular clusters \citep{Helmi2018,Baumgardt2019,Vasiliev2019b} and even to study the internal kinematics of many of these systems: sky-plane rotation \citep{Bianchini2018,Vasiliev2019c,Sollima2019} and PM dispersion and anisotropy \citep{Jindal2019}.
The improved data quality in EDR3 prompted us to reanalyze these properties, and at the same time to explore the fidelity and limitations of EDR3 itself. Since all stars in a given globular cluster have the same true parallax (with negligible spread) and share the same kinematic properties, we may use these datasets (amounting to tens of thousands stars in the richest clusters) to test the reliability of measurements in the \Gaia catalogue and their formal uncertainties.

We apply a number of strict quality filters to select stars that are believed to have reliable astrometric measurements, and use these ``clean'' subsets to determine the properties of the cluster and the foreground populations, and individual membership probabilities for each star, in a mixture modelling procedure detailed in Section~\ref{sec:method}. We then use the selection of high-probability members to assess the statistical and systematic uncertainties on $\varpi$ and $\mu$ in Section~\ref{sec:errors}. After calibrating the recipes for adjusting these uncertainties, we proceed to the analysis of mean PM and parallaxes of clusters in Section~\ref{sec:mean_astrometry}. Namely, we compare the parallaxes with literature distances and find an empirical parallax offset of $\lesssim 0.01$~mas (on top of the zero-point correction applied to each star according to \citealt{Lindegren2021b}) and an additional scatter at the same level. In Section~\ref{sec:orbits} we explore the orbital properties of the entire cluster population, including some of the poorly studied objects with no prior measurements. Then in Section~\ref{sec:internal_kinematics} we analyze the internal kinematics of star clusters: PM dispersions, anisotropy, and rotation signatures. Section~\ref{sec:summary} wraps up.

\section{Method}  \label{sec:method}

We follow the mixture modelling approach to simultaneously determine the cluster membership probability for each star and to infer its properties, in particular, the mean parallax, proper motion, its dispersion, and other structural parameters. Mixture models have an advantage over more traditional cleaning procedures such as iterative $n$-sigma clipping, allowing a statistically rigorous estimation of parameters of the distribution even in cases of low contrast between the cluster and the field populations. Our procedure consists of several steps:
\begin{itemize}
\item First we retrieve all sources from the Gaia archive that are located within a given radius from the cluster centre (this radius is adjusted individually for each cluster and is typically at least a few times larger than its half-light radius; in some cases we increase it even further to obtain a sufficient number of field stars, which are a necessary ingredient in the mixture modelling). At this stage, we do not impose any cuts on sources, other than the requirement to have 5- or 6-parameter astrometric solutions (5p and 6p for short).
\item Then we determine a `clean' subset of sources that have more reliable astrometry, following the recommendations of \citet{Fabricius2021}, but with tighter limits on some parameters. This clean subset excludes sources that have (a) $G<13$, or (b) RUWE $\,> 1.15$, or (c) \texttt{astrometric\_excess\_noise\_sig} $\,>2$, or (d) \texttt{ipd\_gof\_harmonic\_amplitude} $\,> \exp\big[0.18\,(G-33)\big]$, or (e) \texttt{ipd\_frac\_multi\_peak} $\,>2$, or (f) \texttt{visibility\_periods\_used} $\,< 10$, or (g) \texttt{phot\_bp\_rp\_excess\_factor} exceeding the colour-dependent mean trend described by equation~2 and Table~2 of \citet{Riello2021} by more than 3 times the magnitude-dependent scatter given by equation~18 of that paper, or (h) flagged as a duplicated source. If the number of 5p sources satisfying these criteria exceeds 200, we use only these, otherwise take both 5p and 6p sources. The severity of these quality cuts depends on the density of stars: in the central $1-2$ arcmin of some clusters there are virtually no stars satisfying all these criteria. For a few clusters, we slightly relax these thresholds, since otherwise they would have retained too few stars. In addition, we scale the observational uncertainties on $\varpi$ and $\boldsymbol\mu$ by a density-dependent factor discussed in the next section.
\item Then we run a simplified first pass of mixture modelling in the 3d astrometric space only (parallax $\varpi$ and two PM components $\mu_\alpha \equiv ({\rm d}\alpha/{\rm d}t)\cos\delta$, $\mu_\delta \equiv {\rm d}\delta/{\rm d}t$). The distribution of all sources is represented by two (if the total number of stars is below 200) or three Gaussian components, and we use the Extreme Deconvolution (XD) approach \citep{Bovy2011} to determine the parameters (mean values and covariance matrices) of these components, which, after being convolved with observational uncertainties (using the full $3\times3$ covariance matrix provided in the catalogue), best describe the actual distribution of sources. One of these component (usually the narrowest, except NGC~104 and NGC~362, which sit on top of the Small Magellanic Cloud with a lower PM dispersion than the cluster stars) is identified with the cluster, and the remaining one or two components -- with field (usually foreground) stars.
\item Then we run a full mixture model, which differs from the XD model in several aspects. We use the angular distance from the cluster centre $R$ as an additional property of each star (besides $\varpi$ and $\boldsymbol\mu$), and fit the cluster surface density profile $\Sigma(R)$ by a simple Plummer model with the scale radius being a free parameter updated in the course of the MCMC run. Moreover, the intrinsic parallax dispersion of cluster members is set to zero, and the intrinsic PM dispersion is represented by a cubic spline in radius with $2-5$ nodes (depending on the number of stars), with amplitudes that are also varied during the fit. Finally, we allow for spatial variation of the mean PM of cluster stars relative to the mean PM of the entire cluster. We transform the celestial coordinates $\alpha,\delta$, corresponding PM components, and their uncertainty covariance matrix into Cartesian coordinates and velocities on the tangent plane, using the standard orthographic projection (e.g., equation~2 in \citealt{Helmi2018}), and then convert the PM into the rotational (tangential) and radial components. The tangential motion is parametrized by a fixed profile, $\mu_t(R) = \mu_\mathrm{rot}\,2(R/R_0) / \big[ 1 + (R/R_0)^2 \big]$ with $R_0$ equal to the scale length of the Plummer density profile and a free amplitude $\mu_\mathrm{rot}$. The radial PM component of cluster members is assumed to be caused by perspective expansion (an assumption that is tested \textit{a posteriori} in Section~\ref{sec:internal_kinematics}) and hence does not add free parameters. We use the Markov Chain Monte Carlo (MCMC) code \textsc{emcee} \citep{ForemanMackey2013} to explore the parameter space, starting from the astrometric parameters determined by XD and reasonable initial values for the remaining parameters.
\item After the MCMC runs converged, we determine the membership probability for each star (including those that did not pass the initial quality cutoffs), averaging the results of classification scheme from 100 realizations of model parameters drawn from the MCMC chain. The colour-magnitude diagram (CMD) of cluster members is visually inspected to verify the outcome of the mixture model, which did not use the photometric information. The final number of members for each cluster ranges from a few up to more than $10^5$ stars, with typically between 10 and 50\% of stars satisfying all quality filters and hence contributing to the determination of mean properties of the cluster.
\item The mean parallax and PM of the cluster and their uncertainties are taken from the MCMC chain, but these values do not take into account spatially correlated systematic errors, which we include by running an additional postprocessing step, as described in Section~\ref{sec:internal_kinematics}. At this step, we also re-evaluate the internal PM field of cluster members (rotation profile and radial expansion/contraction rate) using a more flexible (spline) parametrization and taking into account systematic errors.
\end{itemize}

This procedure differs from the one used in \citet{Vasiliev2019b,Vasiliev2019c} in several aspects: (a) we typically use two rather than one Gaussian component for the foreground population in the mixture model; (b) we fit the PM dispersion profile simultaneously with other parameters, rather than determining it \textit{a posteriori} from the list of members; (c) we explore the parameter space with MCMC, rather than picking up the single maximum-likelihood solution, which implies propagation of uncertainties in all nuisance parameters (including membership probabilities) into the quantities of interest such as mean PM and its dispersion. The analysis of the entire list of clusters takes a few dozen CPU hours. Similar mixture modelling approaches have been used by \citet{Sollima2020}, who focused on the outer parts of clusters and additionally used photometric information in membership classification, and \citet{Vitral2021}, who employed a fat-tailed Pearson distribution for the field star PM, since it better describes the actual distribution than a single Gaussian (for this reason, we used two Gaussian components for the field population when possible).

\section{Error analysis}  \label{sec:errors}

\begin{figure}
\includegraphics{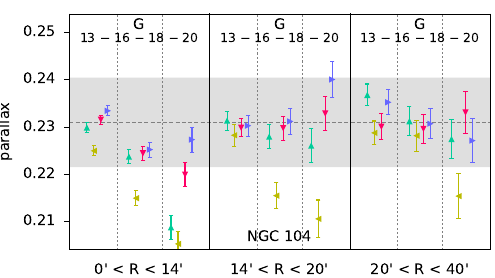}
\includegraphics{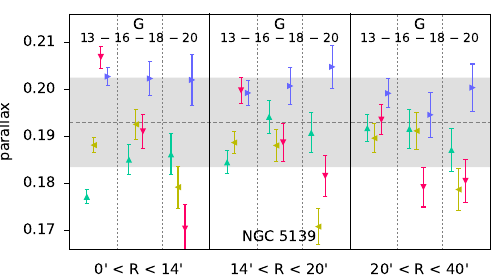}\\[5mm]
\includegraphics{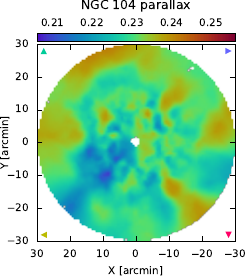}
\includegraphics{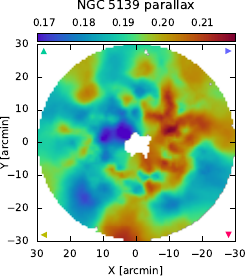}
\caption{Mean parallax of stars in NGC~104 (47~Tuc, top row and bottom left panel) and NGC~5139 ($\omega$ Cen, centre row and bottom right panel) in different magnitude ranges and spatial regions. Three sub-panels in each plot correspond to different radial ranges (centre, intermediate range, and outskirts), and each range is further split into four quadrants, which are shown by different colours and symbols. Each sub-panel is additionally split into three magnitude intervals (bright, intermediate and faint stars). The number of stars contributing to each measurement ranges from $\sim 100-200$ for bright bins to $\sim 2000-3000$ for faint bins. It is clear that the scatter between these split measurements (especially between four quadrants at the same magnitude and radial range) is larger than would be expected from the statistical uncertainties, indicating the presence of an overall systematic error at the level $\sim 0.01$~mas. Gray dotted lines and gray-shaded regions show the mean parallax and its uncertainty, taking into account spatially correlated systematic errors (Equation~\ref{eq:covfncplx}). Bottom row shows the 2d maps of mean parallax (uncertainty-weighted moving average over 200 neighbouring stars brighter than G=18), which also demonstrates spatially correlated residuals at scales of $10-20'$.
}  \label{fig:plx_split}
\end{figure}

\begin{figure*}
\includegraphics{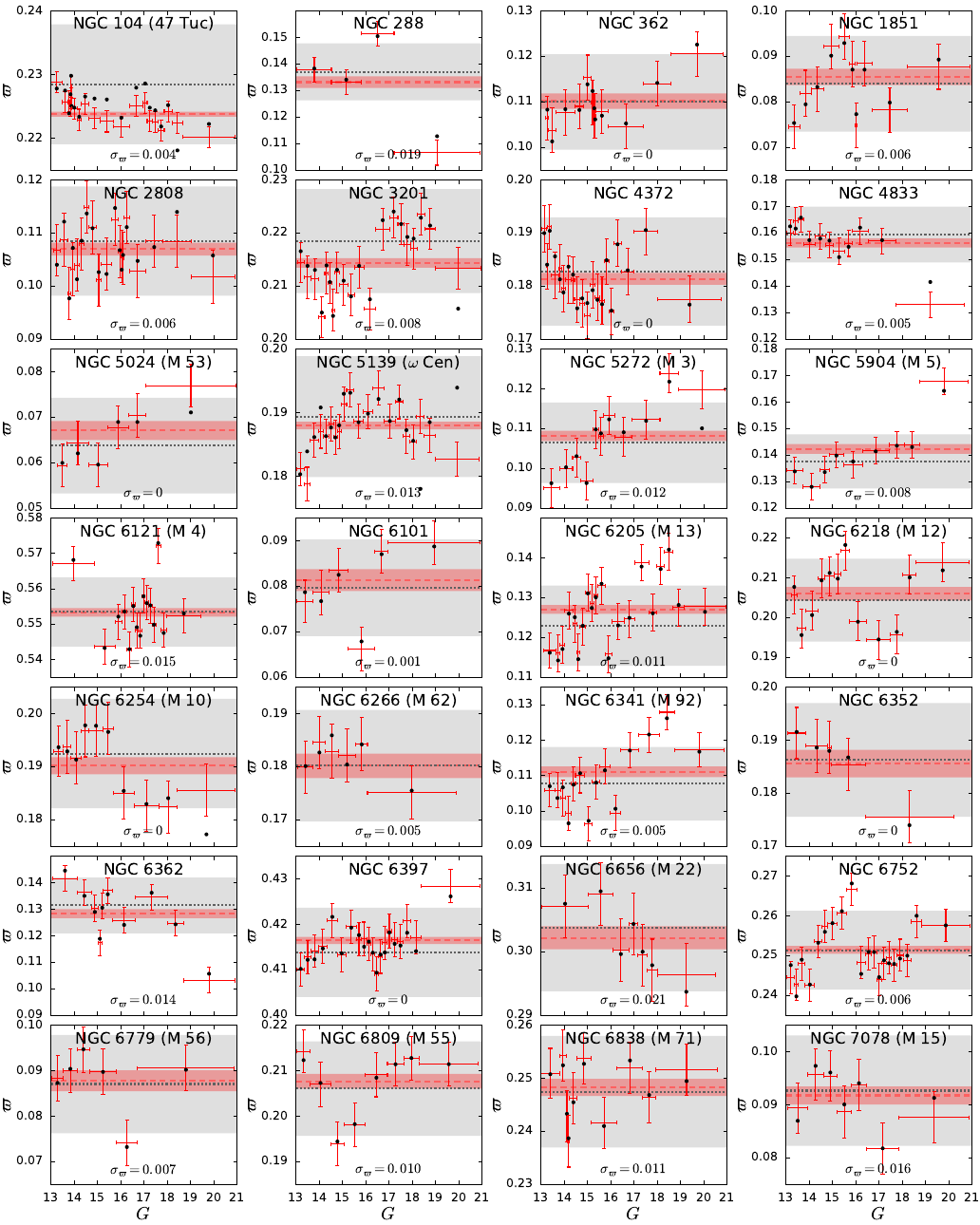}
\caption{Mean parallax of cluster members in different magnitude bins. Red error bars show the value and the statistical uncertainty of conventionally averaged parallaxes, while black dots show the average values taking into account spatially correlated systematic errors (their uncertainties are always at the level $\sim0.01$~mas). Red dashed line and red-shaded region shows the overall mean parallax with statistical uncertainty alone, and black dotted line and gray-shaded region show the overall mean value taking into account correlated systematics. $\sigma_\varpi$ is the intrinsic (statistical uncertainty-deconvolved) spread in parallax values for all cluster stars.
}  \label{fig:plx_mag_trends}
\end{figure*}

\Gaia astrometry has been significantly improved in EDR3, both in terms of lower statistical uncertainties and better calibration. In addition, a larger number of quality criteria are available to filter out stars with possibly unreliable astrometry. Nevertheless, there are still some remaining issues with the statistical and systematic errors, which we explore in this section.
These are grouped into several categories: (1) consistency between averaged values among stars of different magnitudes and spatial regions; (2) underestimated statistical errors; and (3) spatially correlated systematic errors. In these experiments, we use the samples of stars in several of the largest clusters classified as high-confidence members by the mixture model (with $\ge 90$\% probability, although for most stars it actually exceeds 99\%).

The very first question that is natural to ask is whether the average values of parallax and PM are consistent between different subsets of stars split by magnitude or spatial location. Figure~\ref{fig:plx_split} shows the results of this analysis for the two largest clusters with $\sim 50\,000$ members: NGC~104 (47~Tuc) and NGC~5139 ($\omega$~Cen). We split the stars into three magnitude ranges, three radial intervals and four quadrants in each interval. The statistical uncertainties on the mean parallax of stars in each bin are small enough to highlight that the difference between these values exceeds these uncertainties, and is at the level $\sim 0.005-0.01$~mas, depending on the cluster.
Figure~\ref{fig:plx_mag_trends} shows the variation of mean parallax with magnitude for two dozen clusters (without further splitting stars into different spatial regions). Different clusters display a variety of behaviours, ranging from no systematic variation (NGC~2808, NGC~6838) to upward (NGC~5272, NGC~6341) or downward (NGC~6362, NGC~6656) trends with magnitude, or even more complex non-monotonic patterns. This diversity indicates that the parallax zero-point correction suggested by \citet{Lindegren2021b} adequately compensates these variations on average, but not necessarily for each specific region on the sky. The variations of mean parallax in different magnitude bins and the residual scatter of the overall mean parallax (deconvolved from statistical uncertainties of individual stars) are both at the level $0-0.02$~mas depending on the cluster. These variations may not be compensated by the zero-point correction approach, since it relies primarily on sources that are either too sparse and faint (quasars) or concentrated in one region of the sky (stars in the Large Magellanic Cloud -- LMC), so should be considered as an unavoidable additional systematic uncertainty. We now turn to the analysis of its spatial variation.

It is known that the \Gaia astrometry contains spatially correlated systematic errors associated with the scanning law, which vary mostly on the scale $\lesssim 0.5^\circ$, but have some correlations over larger angular scales as well. Although the prominence of these `checkerboard patterns' is significantly reduced in EDR3 compared to DR2 (see, e.g., Figure~14 in \citealt{Lindegren2021a}), they are not completely eliminated. The lower panels of Figure~\ref{fig:plx_split} show the 2d maps of mean parallaxes in NGC~104 and NGC~5139, which indeed fluctuate on a scale of $10-20'$. In the first approximation, these spatially correlated errors in astrometric quantities $\chi\equiv \{\varpi,\mu\}$ can be described by covariance functions $V_{\chi}(\theta) \equiv \overline{\chi_i\,\chi_j}$, which depend on the angular separation $\theta$ between two points $i$, $j$ (measured in degrees).
\citet{Lindegren2021a} estimated these functions in a number of ways, primarily based on the sample of $\sim10^6$ quasars, which are found across the entire sky (except the regions close to the Galactic disc), but are relatively faint. They find $V_\varpi(\theta) \simeq 140\,\mu\mbox{as}^2$ on scales $\sim1^\circ$, and possibly a few times higher in the limit of small separation, although with a large statistical uncertainty limited by the number of close pairs of quasars. On the other hand, LMC stars are brighter and more dense, allowing them to extend the covariance function to smaller separations, which turns out to be $\sim 3\times$ smaller than for quasars. \citet{MaizApellaniz2021} carried out an independent analysis of quasars and LMC stars, but restricted their sample to quasars brighter than $G=19$. They also obtained lower values for $V_\varpi$ in the range $50-80\,\mu\mbox{as}^2$ for separations smaller than a few degrees, and their LMC sample yielded a yet lower $V_\varpi\sim 10-50\,\mu\mbox{as}^2$ for $\theta\lesssim 1^\circ$. All these findings suggest that the spatial correlations are less prominent for brighter stars, but the dependence of the covariance function on magnitude has not yet been quantified (although \citealt{Fardal2021} found that the amplitude of the `checkerboard pattern' in DR2 increases by a factor of a few from bright to faint stars, qualitatively similar to what we see in EDR3).

\begin{figure}
\includegraphics{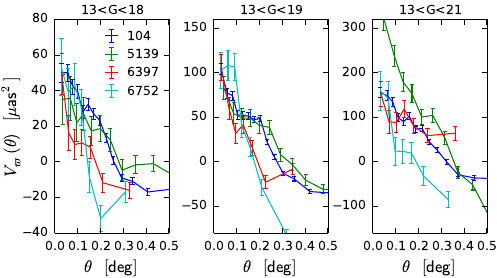}
\caption{Spatial covariance function for parallax as a function of separation between sources $V_\varpi(\theta)$, estimated in four richest clusters in different ranges of magnitudes: from $G=13$ up to $G=18$ (left), $G=19$ (centre) or $G=21$ (right panel), note that each next range includes the previous one. Pairs of stars are binned into $7-15$ distance bins; the uncertainties in each bin are driven by measurement errors, which increase with magnitude. In all panels, the values of $V_\varpi(\theta=0)$ are significantly different from zero, but they increase with the inclusion of fainter stars.
}  \label{fig:covfncplx}
\end{figure}

We explored the spatial covariance of parallaxes and PM in the four richest clusters (NGC~104, 5139, 6397 and 6752), using only the 5p stars that pass all quality criteria, but considering different magnitude ranges. Figure~\ref{fig:covfncplx} shows that the binned covariance functions have a similar behaviour for all clusters, and its limiting value at $\theta=0$ does depend on the magnitude cutoff, increasing towards fainter magnitudes. The value for bright stars ($13<G<18$) is $V_\varpi(0)\simeq 50\,\mu\mbox{as}^2$, consistent with the findings of \citet{MaizApellaniz2021}, and a few times large if we consider all stars. Due to a finite spatial extent of cluster members, the estimated covariance function drops to zero or negative values at separations $\theta\gtrsim 0.2-0.3^\circ$ comparable to the size of the cluster, which does not reflect its true behaviour on large scales (in other words, the systematic error may have spatial correlations on larger scales, but they would be the same for all cluster members and hence simply shift the mean $\varpi$, which is subtracted before computing $V_\varpi$). We thus augment our estimates of $V_\varpi(\theta)$ with the ones from \citet{Lindegren2021a} and \citet{MaizApellaniz2021} for $\theta\gtrsim 0.2^\circ$, while introducing an overall magnitude-dependent normalization factor. Our approximation is
\begin{equation}  \label{eq:covfncplx}
\begin{array}{ll}
V_\varpi(\theta,G) &\!\!\!\!= \big\{ 50/(1 + \theta/0.3^\circ) + 70\,\exp\big[-\theta/30^\circ - (\theta/50^\circ)^2\big]\big\} \\[1mm]
&\!\!\!\!{}\times \mathrm{max}(G-17,1)\;\; \mu\mathrm{as}^2 ,
\end{array}
\end{equation}
where $G$ is the magnitude of the brighter star in the pair. We do not use the same functional form as \citet{MaizApellaniz2021} since the latter is unphysical (a valid covariance function must have a non-negative angular power spectrum, i.e., Legendre integral transform), but instead reproduce the overall trends with a different function.

To conduct a similar analysis for the PM covariance function, we relied on the assumption (tested later in Section~\ref{sec:internal_kinematics}) that the radial component of PM is entirely caused by perspective effects, thus after subtraction of these it should be zero on average. Unfortunately (in this context), unlike parallax, the intrinsic dispersion of PM is significantly larger than measurement errors, thus reducing the statistical significance of the estimated $V_\mu(\theta)$. Nevertheless, we obtain a reasonably consistent limiting value $V_\mu(0)\simeq 400\,[\mu\mbox{as\,yr}^{-1}]^2$ for all clusters and magnitude ranges; combined with the larger-scale trends from \citet{Lindegren2021a}, we approximate
\begin{equation}  \label{eq:covfncpm}
V_\mu(\theta) = 400/(1+\theta/0.3^\circ) + 300\,\exp(-\theta/12^\circ)\;\; [\mu\mathrm{as\,yr}^{-1}]^2 .
\end{equation}

These covariance functions yields a systematic uncertainty $\epsilon_\varpi \simeq 0.011$~mas and $\epsilon_\mu \simeq 0.026$~\masyr in the limit of small separations, which should be viewed as the irreducible systematic floor on the precision of parallax and PM measurement for any compact stellar system. The overall uncertainty (statistical and systematic combined) for a given selection of cluster members is derived using the method described in \citet{Vasiliev2019c}, and may be smaller than these values if a cluster spans more than a fraction of a degree on the sky. \citet{MaizApellaniz2021} discuss a similar method for combining statistical and systematic uncertainties, but in their equations 5--7, the overall error is dominated by the largest values of $V_\varpi(\theta)$, whereas our method essentially sums up the values of the inverse covariance matrix, and therefore puts more emphasis on stars with smallest spatial covariances. In practice, though, the difference should be minor.

\begin{figure*}
\includegraphics{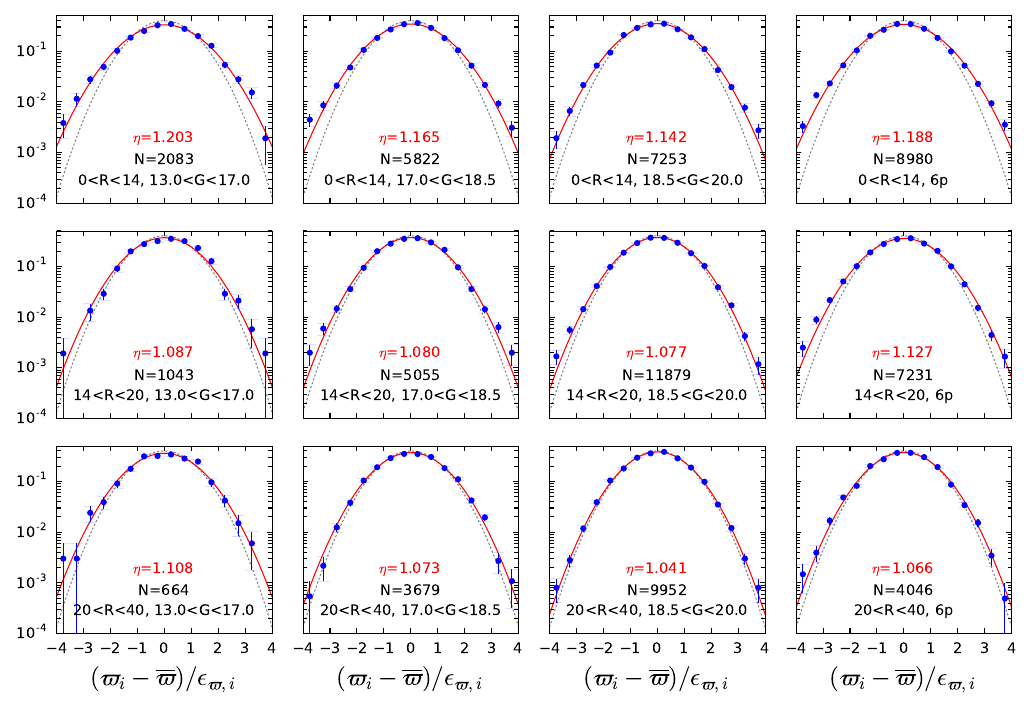}
\caption{Distribution of uncertainty-normalized deviations from the mean parallax for stars in the clean subset in NGC~5139 ($\omega$~Cen). Histograms are split by distance range (from top to bottom row) and magnitude (from left to right, with the last column showing 6p sources, which are typically fainter than $G=20$). If the formal uncertainties were correctly estimated, the histograms should have followed a standard normal distribution (shown by gray dotted lines), however in practice, the distribution is broader by a factor $\eta\simeq 1.05-1.2$, depending on the density of sources (and hence on radius).
}  \label{fig:plx_distribution}
\end{figure*}

\begin{figure*}
\includegraphics{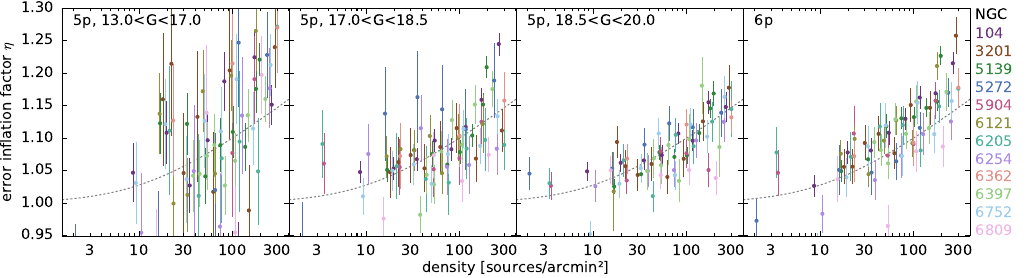}
\caption{Parallax error inflation factor $\eta$ as a function of source density and magnitude, for a dozen clusters with $\ge 5000$ members. For each cluster, we use only the clean subset of stars (5p in the first three panels, and 6p in the last one) split into several equal-number bins in radius; the horizontal axis shows the mean density of stars in each bin, and the vertical axis -- the width of the best-fit Gaussian fitted to the distribution of uncertainty-normalized deviations from the mean parallax for stars in the given bin $(\varpi_i-\overline{\varpi})/\epsilon_{\varpi,i}$, similar to the one shown in Figure~\ref{fig:plx_distribution} for one particular cluster. The scaling factor $\eta$ that the uncertainties should be multiplied by to match the standard normal distribution is generally larger in higher-density central regions; the gray dotted lines show the trend $\eta = (1 + \Sigma/\Sigma_0)^\zeta$ with $\Sigma_0=10$~stars\,arcmin$^{-2}$ and $\zeta=0.04$. The first and the last panel exceed this trend line on average, and a better fit could be obtained by adding in quadrature a systematic error $\epsilon_\mathrm{sys}=0.01$~mas (first panel -- bright stars) or lowering the value of $\Sigma_0$ (last panel -- 6p sources).
}  \label{fig:error_scaling}
\end{figure*}

\begin{figure*}
\includegraphics{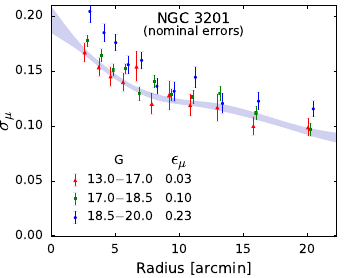}
\includegraphics{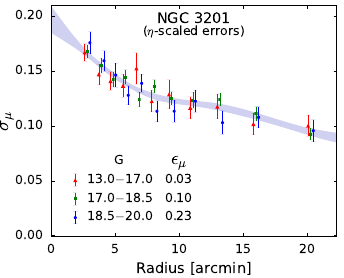}
\includegraphics{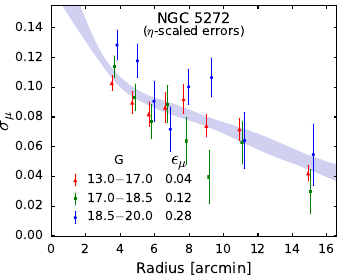}
\caption{PM dispersion profiles for NGC~3201 (left and centre panels) and NGC~5272 (M~3, right panel) as a function of distance, split into several magnitude bins. Left panel: when using nominal measurement uncertainties (the average $\epsilon_\mu$ for each bin is printed in the legend), the deconvolved intrinsic dispersion is higher for fainter stars, indicating that their uncertainties are underestimated. In the remaining two panels, we use the same prescription for the density-dependent error inflation factor $\eta$ as for the parallax, which produces profiles that are in reasonable agreement between different magnitude bins. Shaded areas show the 68\% confidence intervals on $\sigma_\mu(R)$ from the MCMC runs of the full mixture model pipeline, in which this scaling factor was also applied and faint stars have been excluded.
}  \label{fig:pm_dispersion_split}
\end{figure*}

Apart from confirming the systematic error, we also find that the formal statistical uncertainties are somewhat underestimated. This is most easily manifested in the distribution of parallaxes for cluster members: the intrinsic spread of $\varpi$ is negligibly small even for the closest clusters ($\lesssim 10^{-3}$~mas), and hence we should expect the uncertainty-normalized deviations from the mean parallax for each star, $(\varpi_i - \overline{\varpi})/\epsilon_{\varpi,i}$, to follow a standard normal distribution. Figure~\ref{fig:plx_distribution} shows the distribution of these measurements for one of the richest clusters, NGC~5139 ($\omega$ Cen), split into several groups by stellar magnitude and distance from the cluster centre. It is clear that even for the ``clean'' subset of stars that satisfy all quality criteria, the distribution has more prominent tails than a Gaussian, at least in the central regions with high density of stars. We may quantify this effect by assuming that the actual (externally calibrated) measurement uncertainty $\tilde\epsilon_{\varpi}$ is given by the scaled formal uncertainty, summed in quadrature with an additive constant:
\begin{equation}
\epsilon_{\varpi,\rm ext}^2 = \eta^2\,\epsilon_{\varpi}^2 + \epsilon_\mathrm{\varpi,sys}^2 .
\end{equation}

Figure~\ref{fig:error_scaling} shows the error inflation factor $\eta$ estimated in several magnitude bins from a dozen clusters, as a function of source density. Despite some scatter, the overall trends are consistent between all clusters, and indicate that the formal uncertainties should be scaled by $\eta\simeq 1.1-1.15$ in regions with high source density, but $\eta\simeq 1$ in lower-density regions. For brighter stars, $\eta$ is somewhat higher, but one can compensate for this by an additive systematic error $\epsilon_\mathrm{\varpi,sys}\simeq 0-0.02$~mas depending on the cluster (the additive errors for individual clusters are shown on Figure~\ref{fig:plx_mag_trends}). These values are comparable with those reported by \citet[their figures 19, 21]{Fabricius2021}, although they did not study the variation with source density and did not apply all quality criteria that we used here, thus their $\eta$ are somewhat higher on average. \citet{ElBadry2021} and \citet{Zinn2021} find a similarly mild ($\lesssim 1.2$) error inflation factor from analysis of parallaxes of wide binaries and astroseismically calibrated stars, respectively (after applying some quality cuts similar to ours). We also examined the performance of the composite quality filter (``astrometric fidelity``) suggested by \citet{Rybizki2021}, but found that it fails to remove many astrometrically unreliable sources in dense central regions, producing much broader distributions of normalized parallax deviations with $\eta \sim 1.5-2$.

The right panel of Figure~\ref{fig:error_scaling} shows that the scaling factor $\eta$ is slightly higher for 6p sources that pass all other quality filters. We may approximately describe its dependence on source density $\Sigma$ as $\eta = (1 + \Sigma/\Sigma_0)^\zeta$; the values of $\Sigma_0$, $\zeta$ and $\epsilon_\mathrm{sys}$ for various subsets are reported in Table~\ref{tab:error_scaling}.

\begin{table}
\caption{Coefficients for the multiplicative parallax uncertainty scaling factor $\eta = (1 + \Sigma/\Sigma_0)^\zeta$, and the additional systematic error, for different subsets of stars in the range $13<G<20$.
} \label{tab:error_scaling}
\begin{tabular}{llll}
subset & $\zeta$ & $\Sigma_0$ [stars\;arcmin$^{-2}$] & $\epsilon_\mathrm{\varpi,sys}$ [mas] \\
clean, 5p & 0.04 & 10 & 0.01 \\
clean, 6p & 0.04 & 5 & 0.01 \\
non-clean & 0.15 & 20 & 0.04
\end{tabular}
\end{table}

It is natural to expect that the PM uncertainties could be underestimated by a similar factor, but this is more difficult to test empirically, since the intrinsic PM dispersion is non-negligible for most clusters with a sufficiently high number of stars, and is indeed among the free parameters in the mixture model. One possibility is to check whether the error-deconvolved PM dispersion is the same when computed from different magnitude ranges. Figure~\ref{fig:pm_dispersion_split} demonstrates that when using formal uncertainties from the \Gaia catalogue without any correcting factors, the internal PM dispersion appears to be higher for fainter stars, indicating that their uncertainties are likely underestimated. On the other hand, when adopting the same prescription for the PM uncertainty scaling factor $\eta$ as for the parallax, the inferred dispersion is usually consistent between different magnitude ranges. Based on these experiments, we adopted this scaling prescription for the entire mixture model fitting procedure. We also checked that the results are largely insensitive ($\lesssim 0.01$~\masyr difference) to 5\% variations in the adopted value of $\eta\simeq 1.15$ in the highest-density regions (and corresponding changes at lower densities).  

Nevertheless, to avoid possible biases caused by cluster-to-cluster variations of the scaling factor, we conservatively used only those stars from the clean subset that have sufficiently small uncertainties (usually these are the brighter ones).
Namely, we select only stars with $\epsilon_\mu < \kappa\sigma_\mu / (\eta-\eta_0)$, where $\kappa=0.2$ is the tolerance parameter, $\eta(\Sigma)\ge 1$ is the default density-dependent error scaling factor, and we set $\eta_0=0.9$. The idea is that in high-density central regions, $\eta$ is not only higher, but also more uncertain, and thus we reduce the maximum acceptable statistical error of stars in this region to limit the bias from incorrect assumptions about this scaling factor. Obviously, this filter already requires the knowledge of the intrinsic dispersion $\sigma_\mu$, so we apply it iteratively, using the profiles $\sigma_\mu(R)$ obtained in previous runs. All stars in the clean subset that do not pass this filter are still used in the astrometric fit, but are convolved with a fixed (previous) $\sigma_\mu(R)$ profile instead of the one inferred during the fit, so do not bias its properties. If the number of stars with small enough uncertainties is too low ($\lesssim50$), we do not attempt to infer the PM dispersion and instead assume a fixed profile guided by the line-of-sight velocity dispersion, or simply zero for distant or low-mass clusters. A similar cut on statistical uncertainties on the \textit{HST}-derived PM was applied by \citet{Watkins2015a}, who removed all stars with uncertainties larger that $0.5\sigma_\mu$ (this is a more stringent criterion than we use, but their PM uncertainties are typically at the level 0.05~\masyr, which is lower than most \Gaia stars).

\section{Mean parallaxes and proper motions}  \label{sec:mean_astrometry}

\begin{figure*}
\includegraphics{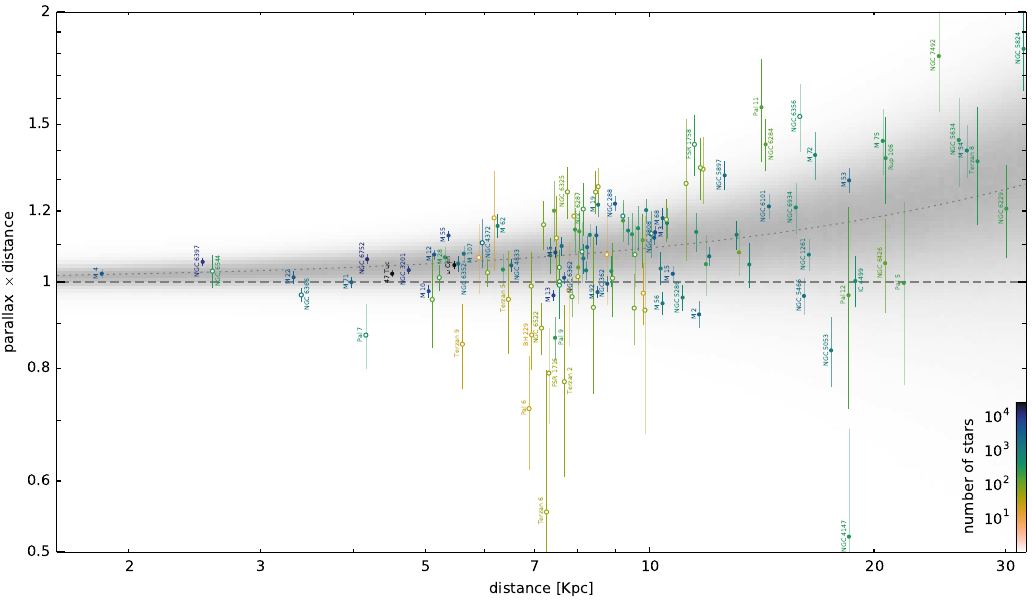}
\caption{Comparison of mean parallaxes $\varpi$ of globular clusters derived in this work with distances $D$ from literature, determined by other methods (dynamical, photometric, etc.). The vertical axis shows the \Gaia parallax multiplied by the distance: values above 1 correspond to the parallax distance being smaller than the literature value, and vice versa. The vertical error bars take into account the statistical uncertainties on both the mean parallax and the distance, but the horizontal error bars for the distance are not displayed; all clusters with $\epsilon_\varpi\,D < 0.2$ are shown. Parallaxes of individual stars that contribute to the mean parallax of a cluster are corrected for the colour- and magnitude-dependent zero-point offset following the recipe from \citep{Lindegren2021b}. Points are coloured according to the number of cluster members, and empty points indicate clusters with higher reddening $E(B-V)\ge 0.5$, for which many photometric distance determination methods may be less reliable, or with fewer than 100 stars. The agreement between \Gaia parallaxes and other distance measurements is fairly good, but at large $D$ the parallaxes seem to be higher on average than $1/D$, and the scatter is larger than the statistical uncertainties. The gray-shaded region shows Monte Carlo samples from a statistical relation (\ref{eq:parallax_and_distance_errors}) reproducing this offset and scatter, with parameters shown in Figure~\ref{fig:parallax_and_distance_errors}.
}  \label{fig:parallax}
\end{figure*}

\begin{figure}
\includegraphics{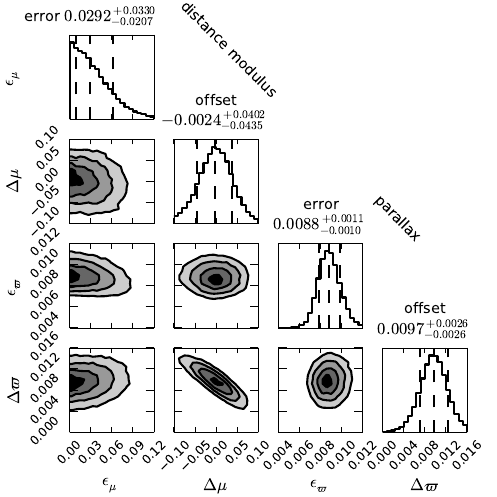}
\caption{Parameters of the statistical relation (\ref{eq:parallax_and_distance_errors}) between \Gaia parallaxes and literature distances, sampled from a Monte Carlo chain. $\epsilon_\mu$ and $\Delta\mu$ are the additional distance modulus error and its offset, both are consistent with zero. $\epsilon_\varpi$ is the systematic uncertainty on \Gaia parallaxes that needs to be added to random uncertainties to make them statistically consistent with distance moduli from the literature, while $\Delta\varpi$ is the parallax offset (positive values indicate that \Gaia parallaxes are on average higher than $1/D$).
}  \label{fig:parallax_and_distance_errors}
\end{figure}

Having established the necessary adjustments and correction factors for the mixture modelling procedure, we now discuss its outcomes pertaining to the global properties of the clusters -- mean parallaxes and PM.

Figure~\ref{fig:parallax} shows the comparison of \Gaia parallaxes with the distances compiled by \citet{Baumgardt2021} from various literature sources. In this plot, the parallaxes and their uncertainties are computed by simple error-weighted averages over member stars (with individual zero-point corrections for each star), without accounting for correlated systematic errors. As expected, the majority of points lie near the line $\varpi\,D=1$, but the deviations from this line are often significantly larger than could be expected from the formal statistical uncertainties alone. Moreover, there is a general tendency of parallax to be slightly larger than $1/D$, illustrated by a large fraction of clusters with $5<D<20$ lying above the $\varpi\,D=1$ line. This may indicate that the parallax zero-point correction suggested by \citet{Lindegren2021b} is slightly overshooting, although it is also possible that there are some general systematic biases in the literature distances. Before exploring this question in detail, we discuss a few specific anomalies.

A number of clusters at low Galactic latitudes have systematically lower parallaxes than implied by the literature distances, but they are in highly-reddenend regions and it is plausible that the CMD-derived distances are biased. In particular, for Pal~6 and Pal~7 (IC~1276), it appears that using the parallax distance and slightly adjusting the extinction coefficient produces a better fit to the CMD than the literature values. 
Parallaxes of NGC~4147 and Terzan~1 are also significantly smaller than $1/D$, and despite rather large uncertainties, the difference is statistically unlikely, possibly indicative of undetected problematic sources in \Gaia. Finally, for some bright, low-extinction clusters such as NGC~6809 (M~55), NGC~6266 (M~62), NGC~288 and NGC~7089 (M~2), the disagreement between CMD-derived and parallax distances exceeds the statistical uncertainty by a factor of few, and cannot be explained other than by the systematic variation of parallax zero-point across the sky.

To explore the possible systematic biases and additional uncertainties, we performed the following analysis. We assume that the deviations of the measured distance moduli\footnote{they are traditionally denoted by the same symbol $\mu$ as the PM.} $\mu_i$ and parallaxes $\varpi_i$ from their true values for $i$-th cluster follow normal distributions:
\begin{equation}  \label{eq:parallax_and_distance_errors}
\begin{array}{l}
\mu_i - \mu_i^{\rm true} \sim 
\mathcal{N} \big( \Delta\mu,\; \epsilon_{\mu,i}^2 + \epsilon_{\mu,\rm sys}^2 \big), \\[1mm]
\varpi_i - 10^{2-0.2\mu_i^{\rm true}} \sim 
\mathcal{N} \big( \Delta\varpi,\; \epsilon_{\varpi,i}^2 + \epsilon_{\varpi,\rm sys}^2 \big),
\end{array}
\end{equation}
where $\epsilon_{\mu,i}$ are formal uncertainties on the distance moduli from the literature, $\epsilon_{\varpi,i}$ are the \Gaia statistical uncertainties, $\epsilon_{\mu,\rm sys}, \epsilon_{\varpi,\rm sys}$ are the additional global systematic uncertainties, and $\Delta\mu, \Delta\varpi$ are global offsets (same for all clusters). For each choice of these four global parameters, we compute the joint likelihood of measured values $\mu_i,\,\varpi_i$ by marginalizing over the unknown true distance modulus:
\begin{equation}  \label{eq:likelihood_distance}
\begin{array}{l}
\displaystyle \mathscr L_i \equiv \int_{-\infty}^{\infty} d\mu_i^{\rm true}\; 
\frac{1}{2\pi\,\sqrt{(\epsilon_{\mu,i}^2 + \epsilon_{\mu,\rm sys}^2)\, (\epsilon_{\varpi,i}^2 + \epsilon_{\varpi,\rm sys}^2)}} \\[6mm]
\displaystyle \times
\exp\left[ 
-\frac{(\mu_i^{\rm true} + \Delta\mu-\mu_i)^2}{\epsilon_{\mu,i}^2 + \epsilon_{\mu,\rm sys}^2}
-\frac{(10^{2-0.2\mu_i^{\rm true}} + \Delta\varpi-\varpi_i)^2}{\epsilon_{\varpi,i}^2 + \epsilon_{\varpi,\rm sys}^2}
\right].
\end{array}
\end{equation}
We then maximize the joint likelihood of all clusters in a Monte Carlo simulation, restricting the sample to clusters with $\ge 100$ stars and low extinction, $E(B-V)<0.5$ (we also tried a number of other quality cuts, but the results were qualitatively similar). Figure~\ref{fig:parallax_and_distance_errors} shows the posterior distribution of these four parameters, and the gray-shaded region in Figure~\ref{fig:parallax} shows the samples from this distribution on the parallax vs.\ distance relation. 
The offset and additional error in distance modulus are consistent with zero, though the uncertainties are at the level of a few$\times10^{-2}$ mag. On the other hand, the offset in \Gaia parallax is $\Delta\varpi \simeq 0.01\pm 0.003$ (that is, zero-point corrected parallaxes are slightly too high), and the additional systematic error in parallax is $\sim 0.01$ -- a very similar value to the one obtained in the previous section from two completely unrelated arguments (spatial covariance function and the variation of mean parallaxes in different magnitude bins). This value likely represents the true limit of the parallax precision in \Gaia EDR3. 

The overcorrection of the parallax zero-point by the \citealt{Lindegren2021b} prescription has been noted in several other studies: for bright stars ($G<13$), \citet{Riess2021} and \citet{Zinn2021} find that the \Gaia parallaxes are overestimated by $\sim0.015$~mas, while \citet{Huang2021} find an overestimation of $\sim 0.009$~mas for stars fainter than $G=14$, with a further position-dependent variation of $\sim 0.01$~mas. Our estimated offset is qualitatively similar to these studies.
Only in special circumstances it may be possible to estimate the zero-point offset more precisely at the location of a given cluster: \citet{Chen2018} combined the sparse quasars with the much more numerous stars of the Small Magellanic Cloud (SMC) to derive corrected parallaxes for NGC~104 (47~Tuc) and NGC~362 from \Gaia DR2. Their values $\varpi_{104}=0.225\pm0.007$, $\varpi_{362}=0.117\pm0.007$ agree well with our measurements (0.232 and 0.114 respectively). Recently \citet{Soltis2021} derived the parallax for NGC~5139 ($\omega$ Cen) from \Gaia EDR3 to be $\varpi_{5139}=0.191\pm0.004$, which, unsurprisingly, agrees with our value 0.193, as well as with a number of other distance estimates from the literature. However, we note that despite a good agreement, this value still carries a systematic uncertainty $\epsilon_\varpi\simeq 0.01$~mas, corresponding to a 5\% distance uncertainty, significantly larger than they optimistically assumed, and reducing the precision of their calibration of cosmological distance indicators. We still need to wait until further \Gaia data releases to bring down the systematic uncertainty to competitive levels.

Comparing our parallax values with the ones derived by \citet{Shao2019} from \Gaia DR2, we find a general agreement within error bars, after correcting for the mean parallax offset in DR2 $\Delta\varpi\simeq -0.03$ (i.e., DR2 parallaxes are smaller on average). Only a few clusters showed a statistically significant disagreement, e.g., NGC~5272 (M~3)  ($\varpi_\mathrm{our}-\varpi_{S19}\simeq0.085$), NGC~6544 ($0.07$), NGC~7099 (M~30) ($0.07$). The parallax values for the few brightest clusters considered in \citet{MaizApellaniz2021} are in complete agreement with our measurements, being derived from the same EDR3 catalogue.
Likewise, the mean PM of clusters in EDR3 are in good agreement with the ones derived by \citet{Helmi2018}, \citet{Baumgardt2019} and \citet{Vasiliev2019b} from \Gaia DR2 data: for 80\% clusters, the total difference in both PM components is within 0.1~\masyr, comparable to the systematic uncertainty of DR2 ($0.066$~\masyr per component). Clusters with the largest PM difference usually have very few stars or are located in dense regions: AM~4 ($\sim 1.7$~\masyr), FSR~1735 ($\sim 1.0$), UKS~1, Terzan~6 (0.8), Terzan~5, Djorg~1 (0.6). Recently \citet{Vitral2021} independently derived mean PM for $\sim100$ globular clusters from \Gaia EDR3, which are in a very good agreement with our results.

The previous analysis involved the mean parallaxes derived using the statistical uncertainties alone. Given the several independent pieces of evidence for the additional systematic error, in the subsequent analysis we use the parallaxes and PM computed with full account for spatially correlated systematic errors, following the method of \citet{Vasiliev2019c}; these values are reported in Table~\ref{tab:catalogue}. The resulting uncertainties on parallax and PM are usually at the level dictated by the covariance functions (\ref{eq:covfncplx}, \ref{eq:covfncpm}), that is, $\epsilon_\varpi\simeq 0.011$~mas, $\epsilon_\mu\simeq 0.026$~\masyr per component; however, they could be slightly smaller for clusters with a large spatial extent, or larger for clusters with too few members. The mean values computed with and without accounting for systematic errors are usually very close, however, there are a few exceptions (mainly for large clusters). For instance, in NGC~104 (47~Tuc) the relatively sparsely populated outer regions have coherently higher parallaxes (lower left panel in Figure~\ref{fig:plx_split}). Since the covariance function drops with distance, the pairs of stars in opposite `corners' of the cluster have a larger contribution to the overall mean value computed while accounting for spatial correlations. On the other hand, the simple average parallax weighted purely with statistical uncertainties of individual stars is dominated by the more numerous stars in the inner part of the cluster, which happen to have somewhat lower parallaxes than the outskirts. As a result, the simple statistical average parallax is lower than the value computed using spatial correlations by $\sim 0.005$~mas (note the offset between the red and the gray lines and points in the upper left panel of Figure~\ref{fig:plx_mag_trends}). However, these offsets in NGC~104 and a few similar clusters are well within the overall systematic uncertainty on the mean parallax or PM.

\begin{figure}
\includegraphics{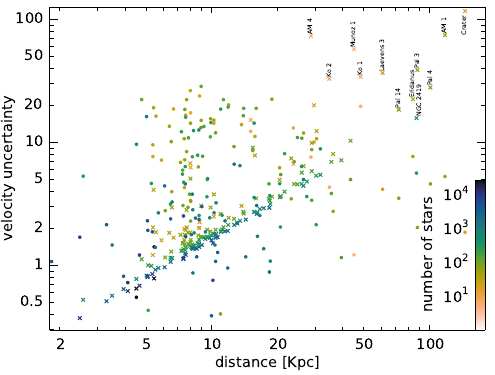}
\caption{Uncertainty in the transverse velocity $v_\bot \equiv \mu\,D$ as a function of heliocentric distance $D$ has two components: distance uncertainty $\mu\,\epsilon_D$ is shown by circles (individual estimates where available, or assuming a rather optimistic 5\% relative error otherwise), and PM uncertainty $\epsilon_\mu\,D$ is shown by crosses (taking into account systematic errors $\epsilon_\mu\simeq 0.025\sqrt{2}$~\masyr, which usually dominate over statistical errors). For most clusters, the first factor is more important. Colours show the number of cluster members with reliable astrometry for each object.
}  \label{fig:velocity_uncertainty}
\end{figure}

The newly derived mean PM are dominated by systematic uncertainties at the level $\sim0.025$~\masyr for most clusters (unless the number of members is below $\sim100$ or the contrast between the cluster and the field stars in the PM space is low, in which case statistical uncertainties may be higher). However, we are ultimately interested not in the value of the PM, but of the transverse physical velocity $v_\bot \equiv \mu\,D$, and the uncertainty in distance usually is the dominant limiting factor in the precision of the velocity, as illustrated in Figure~\ref{fig:velocity_uncertainty}. Overall, the velocity uncertainty is of order $\sim 2-20$~\kms for the majority of clusters, except the most distant ones.

Out of 157 objects in the \citet{Harris1996,Harris2010} catalogue, we could not determine the PM for only a few clusters which are located in highly extincted regions and are not visible to \Gaia: 2MASS--GC01, 2MASS--GC02, GLIMPSE01 and GLIMPSE02. We have also added a number of recently discovered globular cluster candidates to our list: FSR~1716 \citep{Minniti2017}, FSR~1758 \citep{Barba2019}, VVV--CL001 \citep{Minniti2011}, VVV--CL002 \citep{MoniBidin2011}, BH~140 \citep{CantatGaudin2018}, Gran~1 \citep{Gran2019}, Pfleiderer~2 \citep{Ortolani2009}, ESO~93--8 \citep{Bica1999}, Mercer~5 \citep{Mercer2005}, Segue~3 \citep{Belokurov2010}, Ryu~059, Ryu~879 \citep{Ryu2018}, Kim~3 \citep{Kim2016}, Crater / Laevens~1 \citep{Belokurov2014, Laevens2014}, Laevens~3 \citep{Laevens2015}, Mu\~noz 1 \citep{Munoz2012}, BLISS~1 \citep{Mau2019}, bringing the total count to 170. Many of these additional objects are poorly studied and lack line-of-sight velocity measurements, and the nature of them is not well established. We shall see below that a few of these are located at low Galactic latitudes and move within the disc plane, so may well be open rather than globular clusters.

\section{Orbits of globular clusters}  \label{sec:orbits}

\begin{figure*}
\includegraphics{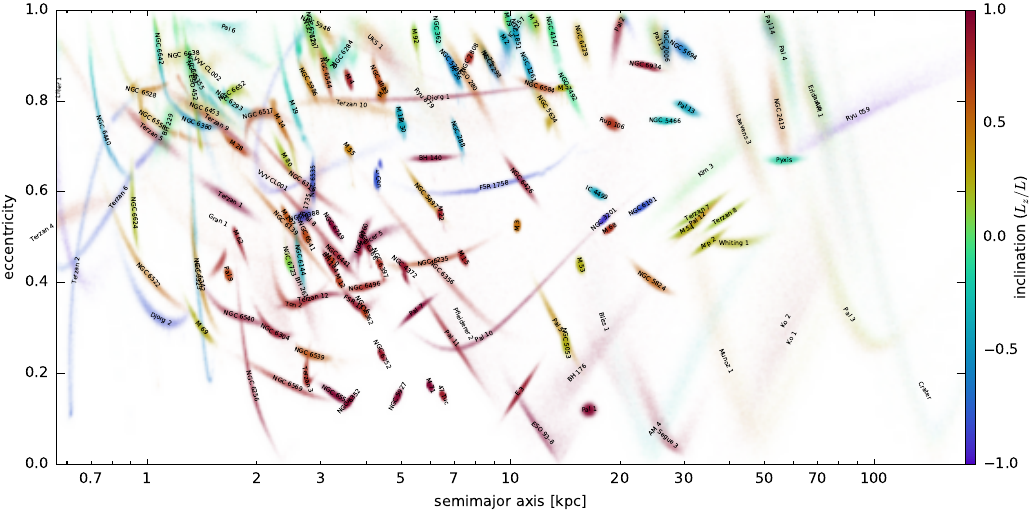}
\includegraphics{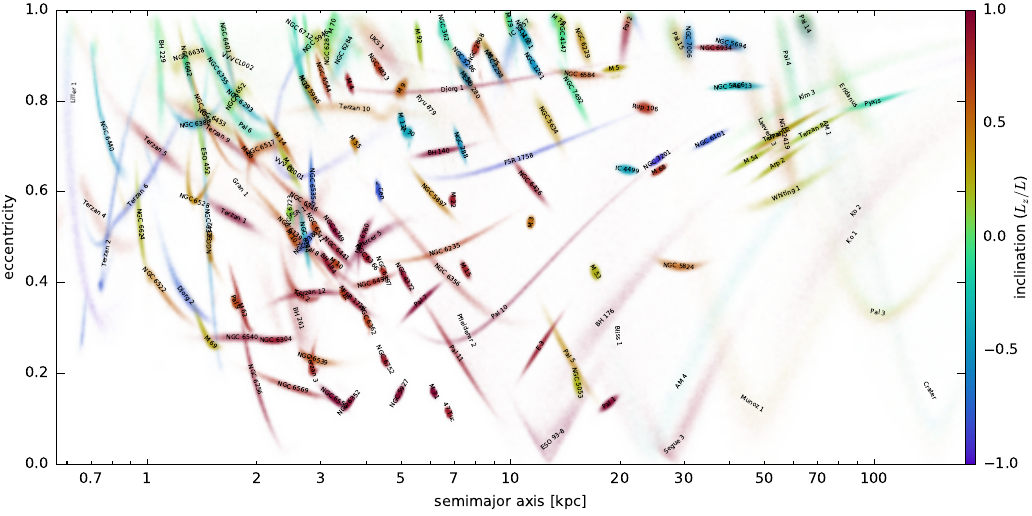}
\caption{Orbital parameters of clusters in the \citet{McMillan2017} best-fit potential (top) and in \citet{Bovy2015} \texttt{MWPotential2014} (bottom). Horizontal axis shows the semimajor axis, vertical -- eccentricity, and colour -- inclination (prograde in red, retrograde in blue, and polar orbits in green). Each cluster is shown by a cloud of points sampled from the measurement uncertainties, which in most cases are dominated by distance uncertainties. We note that the changes in orbit parameters in different Galactic potentials sometimes exceed the measurement uncertainties, especially in the outer part of the Galaxy.
}  \label{fig:orbit_params}
\end{figure*}

Given the full 6d phase-space coordinates of clusters, one may examine their orbital properties or the distribution in the space of integrals of motion: this has become a popular exercise especially after the advent of precise PM from \Gaia DR2 (e.g., \citealt{Binney2017,Myeong2018,Massari2019,Piatti2019,Forbes2020,Kruijssen2020,PerezVillegas2020,Bajkova2020}).
The entire population of clusters can be used to probe the gravitational potential of the Milky Way, using Jeans equations or distribution function-based dynamical models \citep[e.g.,][]{Watkins2019,Posti2019,Vasiliev2019b,Eadie2019}. However, the outer parts of the Galaxy, where this exercise is most useful, are subject to the non-equilibrium distortions caused by the recent passage of the Large Magellanic Cloud (LMC), which perturbs the velocities by as much as few tens \kms and can bias the inference on the gravitational potential \citep{Erkal2020a, Erkal2020b, Cunningham2020, Petersen2021}. Although the effect of the LMC can be accounted for in dynamical models, as illustrated by \citet{Deason2021} in the context of a simple scale-free model for halo stars, we leave a proper treatment of globular cluster dynamics for a future study.

Figure~\ref{fig:orbit_params} shows the orbital properties of the entire population of globular clusters: semimajor axis, eccentricity and inclination, computed in two variants of static Milky Way potentials: \citet{McMillan2017} or \citet{Bovy2015} (these properties remain qualitatively similar if we use other reasonable choices for the potential). Each cluster is shown by a cloud of points representing the measurement uncertainties of its 6d phase-space coordinates, which, as discussed above, are mostly dominated by distance uncertainties (also propagated into the transverse velocity). For the 9 clusters lacking line-of-sight velocity measurements, the missing velocity was drawn from a global distribution function, resulting in a plausible mean value with a large uncertainty $\gtrsim 100$~\kms. An alternative depiction is the rhomboid action-space diagram (e.g., Figure~5 in \citealt{Vasiliev2019b}), which shows the same information in a different projection of the 3d space of integrals of motion. Note however that peri- and apocentres (equivalently, eccentricity and semimajor axis) are computed by numerical orbit integration, whereas actions are computed in the St\"ackel approximation; in practice, this distinction is unimportant.

Objects located in the same region of the plot and having similar colours are close in the 3d integral space and may be physically related, such as the population of globular clusters associated with the Sgr stream. Note that the proximity in the integral space does not imply that the objects line up on the same path on the sky plane, therefore the association with the stream should be examined in the space of observables -- celestial coordinates, distances, PM and line-of-sight velocities. This connection was explored in a number of papers, e.g., \citet{Law2010b}, \citet{Bellazzini2020}, \citet{Arakelyan2020}; however, these studies all relied on the old model of the Sgr stream from \citet{Law2010a}, which was conceived before the more recent observations of its trailing arm, and does not match its features (primarily the distance to the apocentre). When using the model of the Sgr stream from \citet{Vasiliev2021}, which adequately matches all currently available observational constraints, we find that only the following clusters can be unambiguously associated with the Sgr stream: four clusters in the Sgr remnant -- NGC~6715 (M~54, which sits at its centre), Terzan~7, Terzan~8, Arp~2, three clusters in the trailing arm -- Pal~12, Whiting~1 and NGC~2419, and one faint cluster Ko~1, which previously lacked PM measurements (and still has no line-of-sight velocity data), but now coincides with the stream in position, distance and both PM components. Interestingly, it sits in the region where the second wrap of the trailing arm intersects with the second wrap of the leading arm, so is consistent with both interpretations. The line-of-sight velocity is expected to be rather different: 100 to 150~\kms for the trailing arm or $-100$ to 0~\kms for the leading arm. Ko~1 was previously conjectured to belong to Sgr stream by \citet{Paust2014} together with its sibling Ko~2; however, the latter does not match the stream neither in distance nor in PM. A few distant outer halo clusters -- Pal~3, Pal~4 and Crater -- might also be associated with the Sgr debris scattered beyond the apocentre of the trailing arm, but they do not match the model track in at least one dimension (though the model might be unreliable beyond 100~kpc, not being constrained by observational data).

Other conspicuous features in Figure~\ref{fig:orbit_params} include the populations of high-eccentricity clusters with semimajor axes in the range $6-20$~kpc, which are believed to be associated with an ancient merger of a satellite galaxy on a nearly radial orbit \citep{Myeong2018}. Finally, a significant fraction of clusters in the inner part of the Galaxy (with semimajor axes below $6-8$~kpc) have low eccentricity, prograde disc-like orbits (coloured dark red/purple on the plot). Some of the recently discovered objects in the outer part of the Galaxy also share these characteristics. We computed the reflex-corrected PM (assuming that the distance is known to sufficient accuracy), and for several clusters at low Galactic latitudes ($|b|<25^\circ$), the PM component perpendicular to the Galactic plane ($\mu_b^\mathrm{corr}$) is significantly smaller than the parallel component ($\mu_l^\mathrm{corr}$). We may conclude that the orbits of these clusters necessarily stay close to the disc plane, even if the line-of-sight velocity is not known (these cases are marked by italic). This subset of disc-like clusters includes BH~140, BH~176, \textit{BLISS~1}, ESO~93--8, \textit{Ko~2}, Mercer~5, \textit{Pfleiderer~2}, Segue~3. Some of these objects may well be old open clusters rather than globular clusters. The orbit of \textit{Ryu~059} also lies close to the disc plane but is retrograde and highly eccentric (although its line-of-sight velocity is not known, the reflex-corrected sky-plane velocity already exceed 400~\kms), with the estimated apocentre radii exceeding 100~kpc. It is quite plausible that the distance to this cluster is overestimated -- a similar story happened with Djorg~1, for which a 30\% downward distance revision by \citet{Ortolani2019a} made its orbit much less eccentric and more realistic.
Finally, \textit{Kim~3}, Laevens~3, Mu\~noz~1, \textit{Ryu~879} have high-inclination orbits.

As is clear from Figure~\ref{fig:orbit_params}, the orbital parameters often have significant uncertainties, but these are strongly correlated, therefore quoting the confidence intervals on $r_\mathrm{peri/apo}$ or eccentricity makes little sense without a full covariance matrix, or better, the full posterior distribution. Rather than providing these quantities in a tabular form, we provide a Python script\footnote{\url{https://github.com/GalacticDynamics-Oxford/GaiaTools}} for computing orbits in any given potential, sampling from the possible range of initial conditions for each cluster; the integrations are carried out with the \textsc{Agama} library for galactic dynamics \citep{Vasiliev2019a}.

\section{Internal kinematics}  \label{sec:internal_kinematics}

Figure~\ref{fig:profiles} in the Appendix shows the derived PM dispersion and rotation profiles for more than $100$ clusters that have a sufficiently large number of stars and are not too distant, comparing our results with other studies.

\begin{figure}
\includegraphics{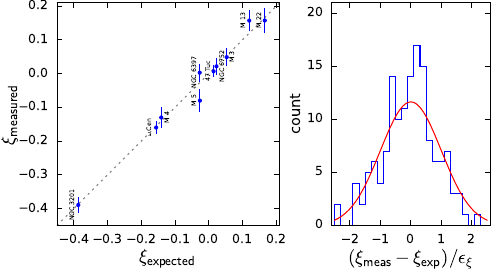}
\caption{Slope of the radial PM component $\xi \equiv \mu_R / R$ (in units of \masyr per degree). 
Left panel shows the measured values for 10 clusters with uncertainties $\epsilon_\xi$ smaller than 0.05, which agree fairly well with values expected from perspective effects. Right panel shows the distribution of deviation of measured values from expectations, normalized by measurement uncertainties $\epsilon_\xi$, for $\sim150$ clusters with at least 100 stars. The red curve shows the standard normal distribution, which adequately describes the actual histogram of deviations (though the latter has a slight excess of objects around zero, for which the uncertainties  $\epsilon_\xi$ might be overestimated).
} \label{fig:pmrad}
\end{figure}

\begin{figure*}
\includegraphics{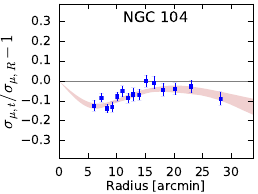}%
\includegraphics{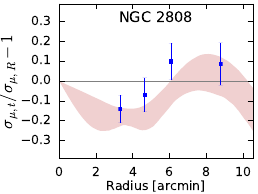}%
\includegraphics{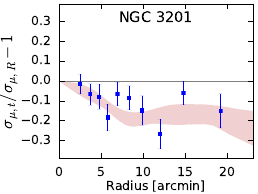}%
\includegraphics{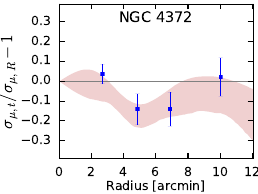}
\includegraphics{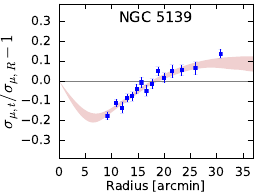}%
\includegraphics{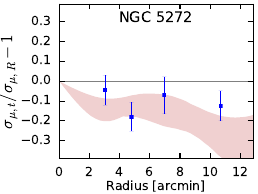}%
\includegraphics{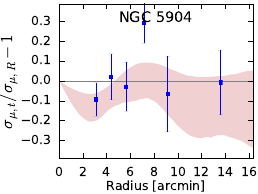}%
\includegraphics{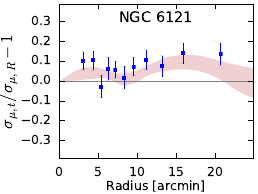}
\includegraphics{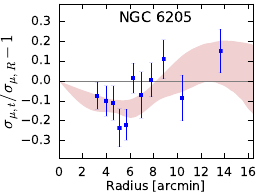}%
\includegraphics{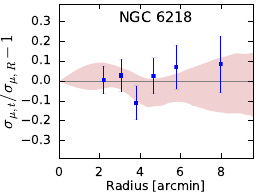}%
\includegraphics{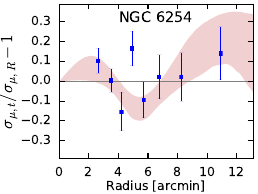}%
\includegraphics{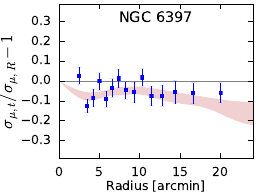}
\includegraphics{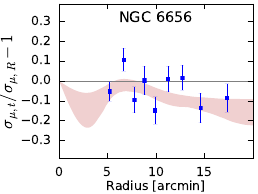}%
\includegraphics{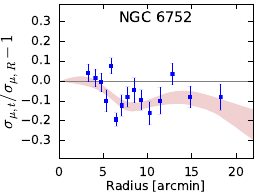}%
\includegraphics{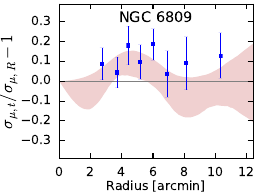}%
\includegraphics{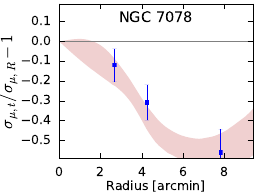}
\caption{PM anisotropy profiles for the richest clusters. Shown is the ratio $\sigma_{\mu,t}/\sigma_{\mu,R}-1$ as a function of radius, represented either as a spline (shaded regions show the 68\% confidence intervals) or split into bins with $\sim500$ stars per bin (except the two largest clusters); only the stars from the clean subset with small uncertainties are used. The profiles are rather diverse: half of the clusters (NGC~104, NGC~2808, NGC~3201, NGC~4372, NGC~5272, NGC~6205, NGC~6397, NGC~6752 and especially NGC~7078) are radially anisotropic, some (NGC~6121, NGC~6809) show weak tangential bias, NGC~5139 transitions from radial to tangential anisotropy, and remaining objects are consistent with isotropy.
}  \label{fig:anisotropy}
\end{figure*}

We explored the sky-plane rotation signatures in all clusters, using the method detailed in the Appendix~A6 of \citet{Vasiliev2019c}, which takes into account spatially correlated systematic errors. Namely, we fitted a general linear model with $N+3$ free parameters to the PM field of stars with high membership probability, where the free parameters are the two components of the mean PM, the slope $\xi$ of the radial PM component ($\mu_R = \xi\,R$), and $N$ amplitudes of a B-spline representation of the tangential PM component $\mu_t$, with $N=2-4$ depending on the number of stars. As discussed in that paper, it is not possible to combine the analysis of spatially correlated systematic errors and probabilistic membership, so we used it as a post-processing step after the main MCMC run, and accounted for uncertain membership by considering 16 realizations of the subset of cluster members, selecting stars in proportion to their membership probability. The PM dispersion profile was kept fixed, as the previous analysis indicated that it is little affected by the spatial correlations (unlike the mean PM field). 

The radial PM component is expected to be caused entirely due to perspective contraction or expansion: for a cluster at a distance $D$ moving with line-of-sight velocity $v_\mathrm{los}$, the expected $\mu_R$ at an angular distance $R$ (measured in degrees) is $\xi\,R$, with $\xi_\mathrm{expected} = -v_\mathrm{los} / D \times (\pi/180^\circ/4.74)$.
Figure~\ref{fig:pmrad}, left panel, compares the measured and expected values of $\xi$ for 10 clusters with small measurement uncertainties $\epsilon_\xi$, demonstrating a very good agreement. The cluster with the largest amplitude of the perspective expansion is NGC~3201, for which the value of $\xi$ is measured with $\lesssim6\%$ relative error; for NGC~5139 ($\omega$ Cen) the relative error is $\sim12\%$. The right panel of the same figure shows the histogram of differences between measured and expected values, normalized by the measurement uncertainties, which closely follows the standard normal distribution (though with an excess of points around zero, for which the uncertainties $\epsilon_\xi$ might be overestimated). 
We stress that these uncertainties take into account spatially correlated systematic errors: if we use only statistical errors, $\epsilon_\xi$ would be considerably smaller for rich clusters, and deviations between measured and expected $\xi$ would exceed $(3-4)\epsilon_\xi$ for quite a few objects. This exercise validates our approach for treating correlated systematic errors and gives credence to the similar analysis of rotation signatures in the PM.

\begin{table}
\caption{Rotation signatures in star clusters detected in this work (last column)
and in some previous studies based on \Gaia data:
\citet{Helmi2018}, \citet{Bianchini2018}, \citet{Vasiliev2019c}, \citet{Sollima2019}.
``$+$'' indicates a firm detection, ``?'' -- a tentative detection, and ``$-$'' stands for no clear signature.
} \label{tab:rotation}
\begin{tabular}{lccccc}
Cluster  & G18 & B18 & V19 & S19 & \makebox[7mm][l]{this work} \\
NGC 104  & $+$ & $+$ & $+$ & $+$ & $+$\\
NGC 1904 &     & $-$ &     &  ?  & $+$\\
NGC 3201 &     &  ?  & $-$ &  ?  & $+$\\
NGC 4372 &     & $+$ &  ?  &  ?  & $+$\\
NGC 5139 & $+$ & $+$ & $+$ & $+$ & $+$\\
NGC 5272 & $+$ & $+$ &  ?  &  ?  & $+$\\
NGC 5904 & $+$ & $+$ & $+$ & $+$ & $+$\\
NGC 5986 &     & $-$ & $-$ &  ?  &  ? \\
NGC 6139 &     &     &     &     & $+$\\
NGC 6218 &     & $-$ & $-$ &  ?  & $+$\\
NGC 6266 &     &  ?  & $+$ & $+$ & $+$\\
NGC 6273 &     & $+$ & $+$ & $+$ & $+$\\
NGC 6333 &     &     &     &     &  ? \\
NGC 6341 &     & $-$ & $-$ &  ?  & $+$\\
NGC 6388 &     & $-$ & $-$ & $-$ &  ? \\
NGC 6402 &     &  ?  & $-$ &  ?  & $+$\\
NGC 6539 &     &  ?  & $-$ &  ?  &  ? \\
NGC 6656 & $+$ & $+$ & $+$ & $+$ & $+$\\
NGC 6715 &     &     & $-$ & $-$ &  ? \\
NGC 6752 & $+$ & $+$ &  ?  &  ?  & $+$\\
NGC 6809 & $+$ & $+$ &  ?  &  ?  &  ? \\
NGC 7078 & $+$ & $+$ & $+$ & $+$ & $+$\\
NGC 7089 &     & $+$ & $+$ & $+$ & $+$\\
\end{tabular}
\end{table}

The uncertainty in the PM rotation profile is typically at the level $0.01-0.015$~\masyr for sufficiently rich clusters, being dominated by systematic errors. We detect unambiguous rotation (at more than $3\sigma$ level) in 17 clusters, with further 6 showing indicative signatures exceeding $0.03$~\masyr at $\sim 2\sigma$ level; they are listed in Table~\ref{tab:rotation}. In most cases, our findings agree with previous studies based on \Gaia DR2 (also shown in the table). The new additions are NGC~6139, in which we find a rather prominent rotation signature despite a lack of it in DR2; NGC~6333 (M~9) and NGC~6388, in which the signal in EDR3 is stronger than in DR2 but still not unambiguous; and NGC~6715 (M~54), in which \citet{AlfaroCuello2020} detected rotation in line-of-sight velocities (although this cluster is a rather special case, sitting in the centre of the Sgr dwarf galaxy, and our kinematic analysis does not separate it from the stars of the galaxy itself). On the other hand, we do not detect significant rotation in several clusters examined and found rotating by \citet{Sollima2019}: NGC~2808, NGC~6205 (M~13), NGC~6397, NGC~6541, NGC~6553, NGC~6626 (M~62): that study considered both PM and line-of-sight velocities, and all these clusters have inclination angles exceeding 60$^\circ$, i.e., the rotation signal is mostly seen in the line-of-sight velocity field (though we do see weak signatures in the PM field of the last two objects). We also excluded Terzan~5 due to a small number of stars satisfying our quality cuts.

Turning to the analysis of PM dispersion profiles, we compare them with the profiles derived from \Gaia DR2 by \citet{Vasiliev2019c} and \citet{Baumgardt2019}, finding generally a good agreement, with the present study having somewhat smaller uncertainties due to improvements in \Gaia astrometry. In addition, the inferred PM dispersion profiles can be compared with the \textit{HST}-derived PM dispersions in the central parts of 22 clusters studied in \citet{Watkins2015a} and 9 clusters studied in \citet{Cohen2021}. Unfortunately, due to the strict quality cutoffs adopted in the present study, there is very little (if any) spatial overlap between \Gaia and \textit{HST} measurements from \citet{Watkins2015a}, but in general, the PM dispersion profiles agree remarkably well.
Some discrepancy is seen in NGC~5139 ($\omega$ Cen), where our PM dispersion profile is lower (although there are almost no stars in our clean subset to anchor it in the crowded central parts), NGC~6341 (M~92), where \Gaia $\sigma_\mu$ is slightly higher, NGC~6535, where both profiles have large uncertainties but \Gaia is lower, and NGC~6681 (M~70), where the \Gaia $\sigma_\mu$ profile is significantly lower. The last case is the most puzzling discrepancy, which appears to be robust against variations in the assumed error inflation factor $\eta$ or various quality filters.
The 9 clusters with \textit{HST} measurements from \citet{Cohen2021} have a larger spatial extent and agree very well with \Gaia in all cases except NGC~6355 and NGC~6401, where \Gaia $\sigma_\mu$ is slightly higher, and NGC~6380, where \Gaia dispersion is $\sim15\%$ lower.

The PM dispersion profiles can also be compared with line-of-sight velocity dispersion profiles from various spectroscopic studies; here we use three such datasets -- \citet{Kamann2018} used MUSE IFU in central regions of some clusters, while \citet{Ferraro2018} and \citet{Baumgardt2019} provide wide-field coverage. The conversion from $\sigma_\mu$ to $\sigma_\mathrm{LOS}$ involves distance, and is one of the methods for its determination. Using the average distances from the literature, we generally get good agreement between PM and line-of-sight dispersions, but in some cases the discrepancy is significant and may indicate some problems in either dataset, or alternatively, calls for a revision of the distance. For well-populated clusters, the discrepancies are usually less than 10\%. The most notable outliers are NGC~1904, NGC~5272 (M~3), NGC~6388, where \Gaia PM is $\sim10-15\%$ higher for the adopted distances; NGC~6304, NGC~6553, NGC~6626, NGC~6779 (M~56), where \Gaia PM is $\sim15\%$ lower; and NGC~6681 (M~70), which is problematic as described above.

Although we used isotropic PM dispersion in the mixture model by default, in the richest clusters we were able to explore the anisotropy by fitting the radial and tangential PM dispersions separately. Figure~\ref{fig:anisotropy} shows the radial profiles of PM dispersion anisotropy, $\sigma_t/\sigma_R-1$, for 16 clusters with sufficient number of stars and satisfying the consistency checks that the PM dispersion and anisotropy agree between stars in different magnitude ranges. There is a considerable diversity in the anisotropy profiles: half of these clusters have predominantly radially anisotropic PM dispersion, a few others have tangential anisotropy, and NGC~5139 ($\omega$~Cen) transitions from being radially anisotropic in the inner part to tangentially anisotropic in the outskirts. The difference between $\sigma_R$ and $\sigma_t$ is typically at the level $10-20$\%, except NGC~7078 (M~15) -- a core-collapsed cluster with a rather strong radial anisotropy in the outer part. Reassuringly, the line-of-sight velocity dispersion outside the core of M~15 matches well the tangential component of PM dispersion (since both mainly reflect the tangential component of 3d velocity), while the radial component of PM dispersion is noticeably higher (another reason why the line-of-sight velocity should \textit{never} be called ``radial velocity''!).

PM anisotropy was explored by \citet{Watkins2015a} in the central parts of 22 clusters observed by \textit{HST}, who found deviations from isotropy at a few per cent level. However, as discussed above, there is almost no spatial overlap between \textit{HST} and the clean \Gaia subset. Our findings can be more directly compared to \citet{Jindal2019}, who explored PM anisotropy in 10 clusters using \Gaia DR2. Our results agree for all clusters except NGC~6656 (M~22), which they found to be radially anisotropic but we do not see a strong evidence for this, and NGC~6397, which appeared to be isotropic in their analysis but weakly radial in our study. The radial anisotropy in NGC~3201 has also been detected by \citet{Bianchini2019} using \Gaia DR2, consistent with our measurement.

\section{Summary and discussion}  \label{sec:summary}

\Gaia EDR3 is a quantitative rather than qualitative improvement upon the revolutionary DR2 catalogue, yet its precision is materially better. In the first part of this study, we scrutinized the fidelity and accuracy of \Gaia astrometry with a variety of methods, using the data from a dozen richest globular clusters, after applying a number of stringent filters to remove possibly unreliable sources.
\begin{itemize}
\item Formal statistical uncertainties on parallax and PM adequately describe the actual errors in regions with low stellar density, but appear to be underestimated by $10-20\%$ in higher-density regions; Table~\ref{tab:error_scaling} provides the suggested correction factors.
\item The parallax zero-point correction proposed by \citet{Lindegren2021b} might be too large (overcorrecting) by $\sim 0.01\pm0.003$~mas.
\item Spatially correlated systematic errors in parallax and PM are considerably lower than in DR2: for bright stars, the residual systematic error in $\varpi$ is at the level $0.01$~mas (a fourfold improvement), while for stars fainter than $G=18$ it may be twice higher. The systematic uncertainty in PM is $\sim0.025$~\masyr, or $2.5\times$ better than in DR2.
\end{itemize}

In the second part, we re-examined the kinematics of almost all Milky Way globular clusters, derived the mean parallaxes and PM, galactic orbits, and analyzed the internal rotation, dispersion and anisotropy profiles of sufficiently rich clusters. While the improvements in precision for mean PM with respect to studies based on DR2 (\citealt{Helmi2018}, \citealt{Baumgardt2019}, \citealt{Vasiliev2019b}) is substantial, the precision of the phase-space coordinates is typically limited by distance rather than PM uncertainties. \Gaia parallaxes will eventually deliver the most precise distance estimates for nearby clusters, but at present, the systematic uncertainty at the level $0.01$~mas limits their usefulness in practice. On the other hand, the internal kinematics (PM dispersions and rotation signatures) have also improved considerably, and agree well with various independent estimates. We find evidence of rotation in more than 20 clusters, and measure the PM dispersion profiles in more than a hundred systems, down to the level $0.05$~\masyr, i.e. at least $2\times$ better than in DR2. These data can be used to improve dynamical models of clusters and provide independent distance estimates, which are examined in \citet{Baumgardt2021}. 

\section*{Acknowledgements}
EV acknowledges support from STFC via the Consolidated grant to the Institute of Astronomy. We thank A.Riess and G.Gontcharov for valuable discussions.
This work has made use of data from the European Space Agency (ESA) mission \Gaia (\url{https://www.cosmos.esa.int/gaia}), processed by the \Gaia Data Processing and Analysis Consortium (DPAC, \url{https://www.cosmos.esa.int/web/gaia/dpac/consortium}). Funding for the DPAC has been provided by national institutions, in particular the institutions participating in the \Gaia Multilateral Agreement.

\section*{Data availability}

We provide the catalogues of all \Gaia sources in the field of each cluster with their membership probabilities and the tables of radial profiles of PM dispersion and rotation amplitudes, available at \url{https://zenodo.org/record/4549397}, as well as the summary table of mean parallaxes and PM and scripts for computing cluster orbits in a given potential, available at \url{https://github.com/GalacticDynamics-Oxford/GaiaTools}.


\appendix
\section{Additional plots and tables}
\begin{figure*}
\includegraphics{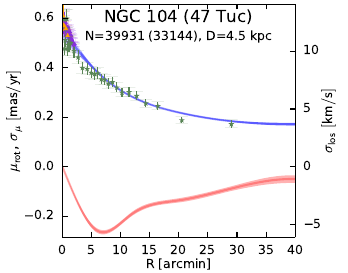}%
\includegraphics{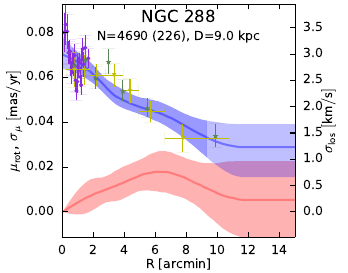}%
\includegraphics{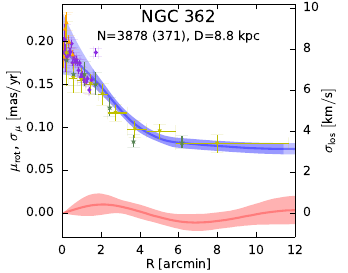}
\includegraphics{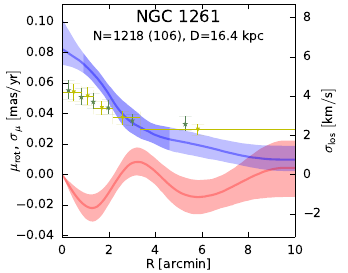}%
\includegraphics{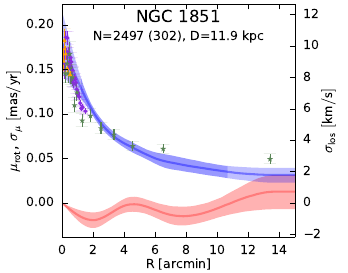}%
\includegraphics{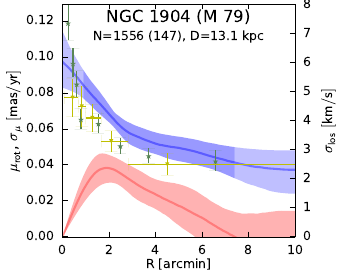}
\includegraphics{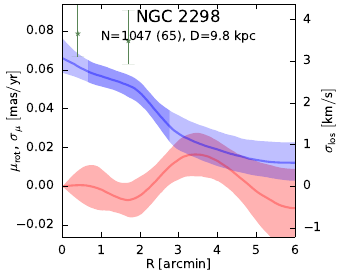}%
\includegraphics{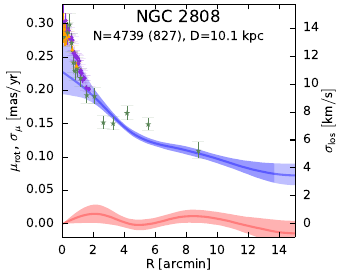}%
\includegraphics{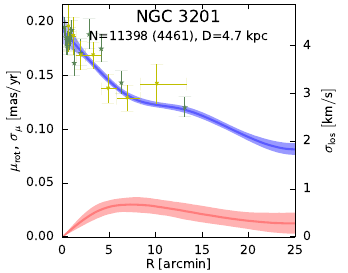}
\includegraphics{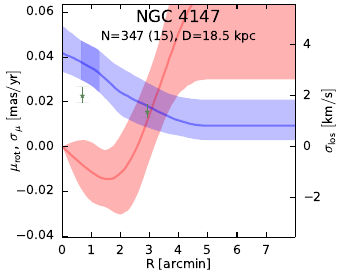}%
\includegraphics{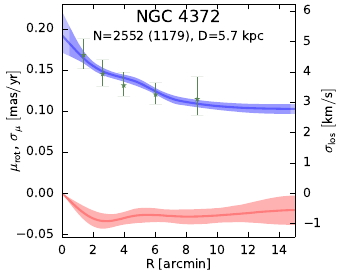}%
\includegraphics{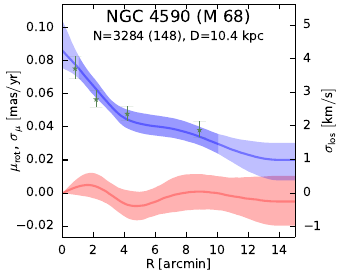}
\caption{
Kinematic profiles of Milky Way globular clusters derived in this work and in other studies.\protect\\
Blue and red solid lines show the radial profiles of internal PM dispersion $\sigma_\mu$
and mean rotation $\mu_\mathrm{rot}$; shaded bands depict 68\% confidence intervals taking into account
systematic errors.
Violet diamonds show the PM dispersion profiles from \textit{HST} \citep{Watkins2015a,Cohen2021};
orange upward triangles, yellow downward triangles and greenish-gray stars -- line-of-sight velocity
dispersions from \citet{Kamann2018}, \citet{Ferraro2018} and \citet{Baumgardt2018}, correspondingly.
$N$ refers to the number of cluster members with good astrometry, and the number in brackets -- to the
number of stars with small enough uncertainties to be used in the measurement of PM dispersion;
the light-shaded part of the PM dispersion profile shows the ranges of radii containing less than 5 stars
at both ends.
\textit{(Continued on next page)}
}  \label{fig:profiles}
\end{figure*}

\begin{figure*}
\includegraphics{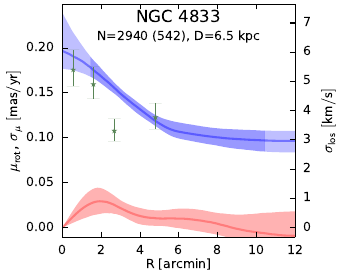}%
\includegraphics{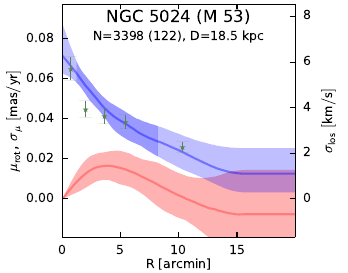}%
\includegraphics{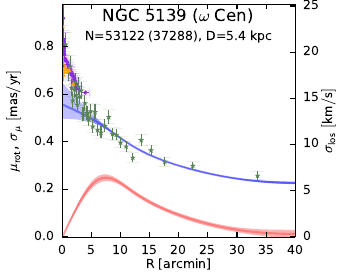}
\includegraphics{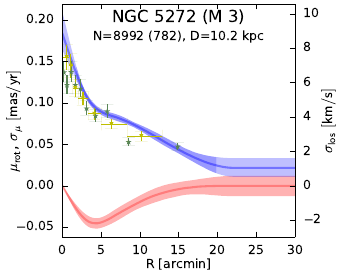}%
\includegraphics{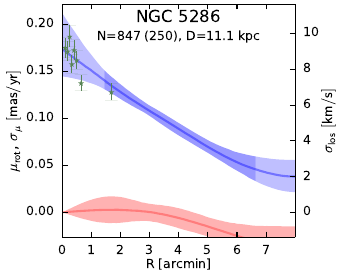}%
\includegraphics{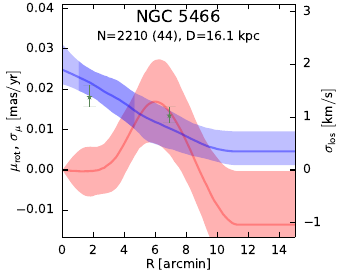}
\includegraphics{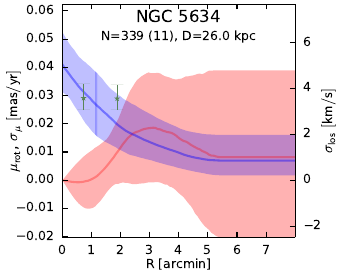}%
\includegraphics{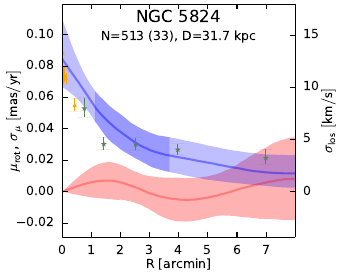}%
\includegraphics{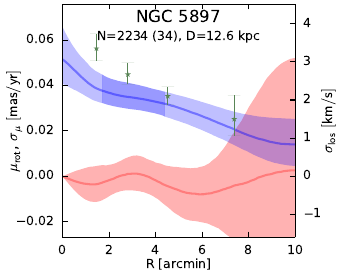}
\includegraphics{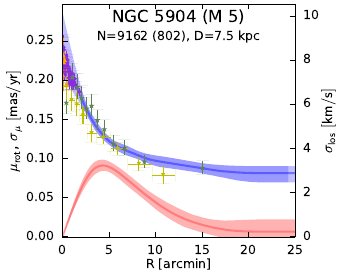}%
\includegraphics{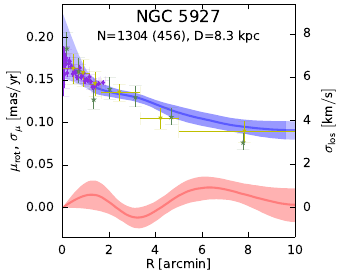}%
\includegraphics{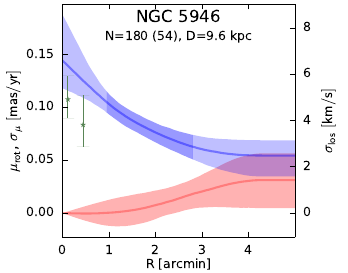}
\includegraphics{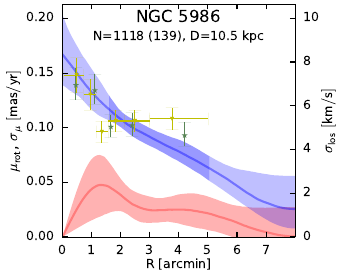}%
\includegraphics{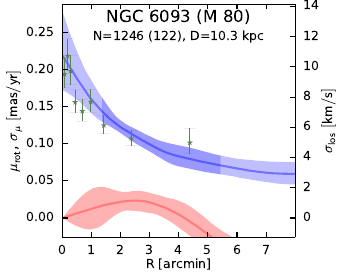}%
\includegraphics{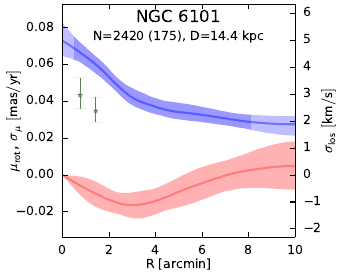}
\contcaption{}
\end{figure*}

\begin{figure*}
\includegraphics{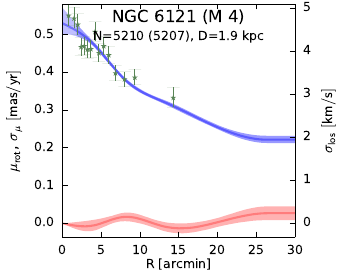}%
\includegraphics{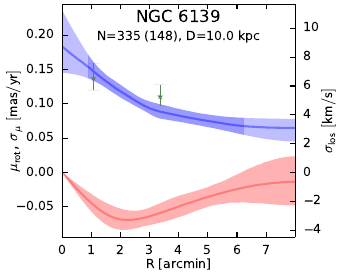}%
\includegraphics{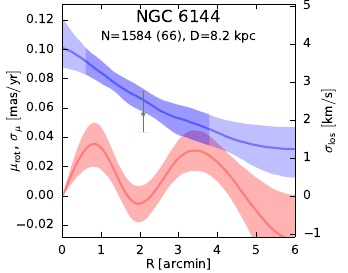}
\includegraphics{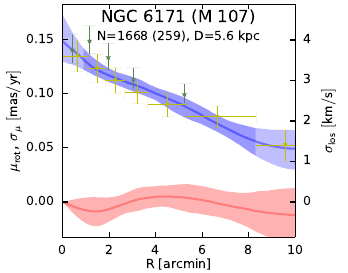}%
\includegraphics{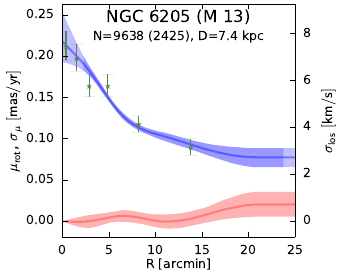}%
\includegraphics{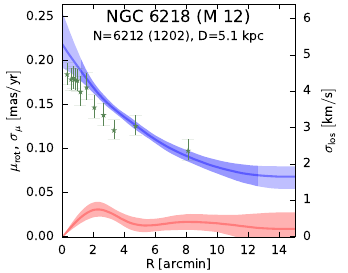}
\includegraphics{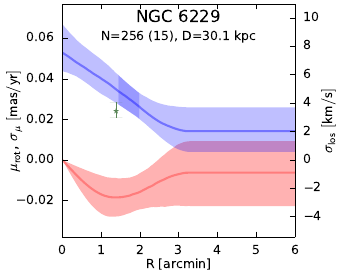}%
\includegraphics{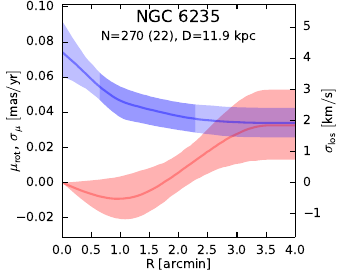}%
\includegraphics{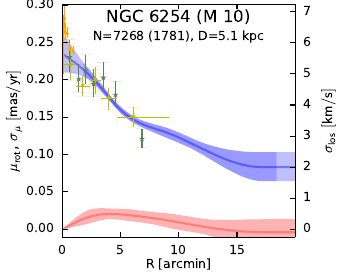}
\includegraphics{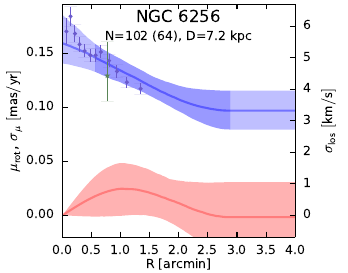}%
\includegraphics{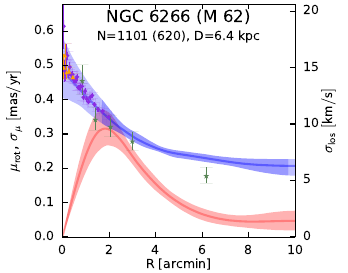}%
\includegraphics{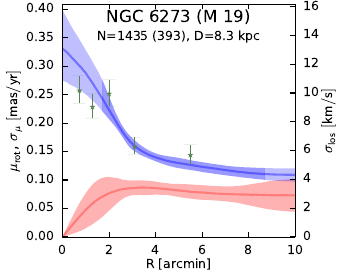}
\includegraphics{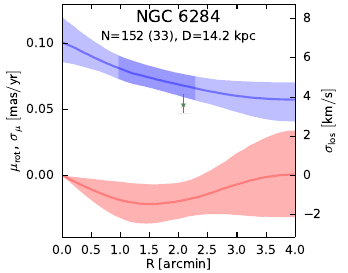}%
\includegraphics{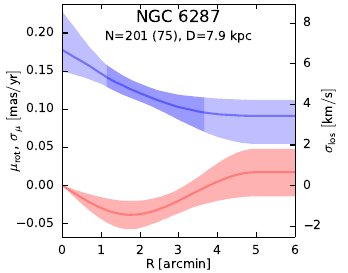}%
\includegraphics{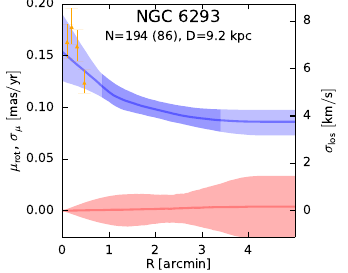}
\contcaption{}
\end{figure*}

\begin{figure*}
\includegraphics{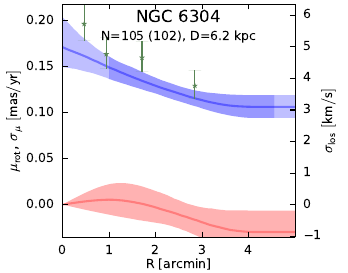}%
\includegraphics{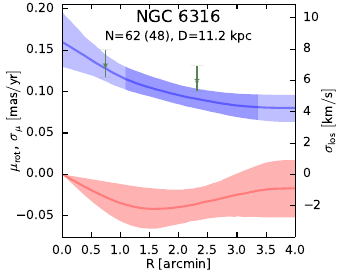}%
\includegraphics{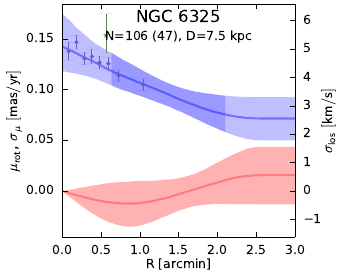}
\includegraphics{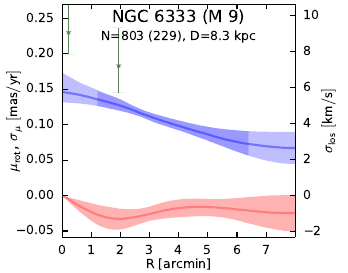}%
\includegraphics{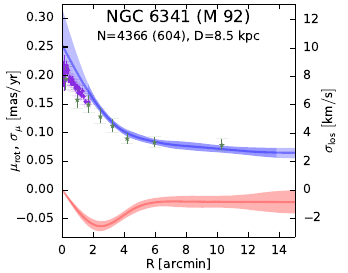}%
\includegraphics{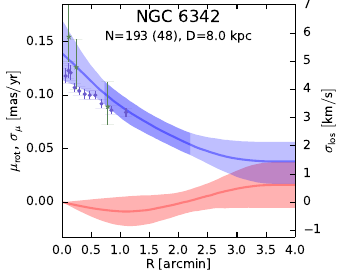}
\includegraphics{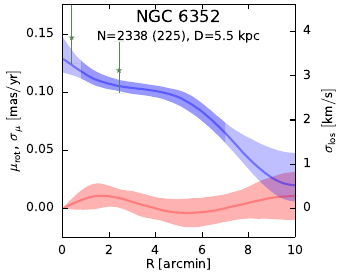}%
\includegraphics{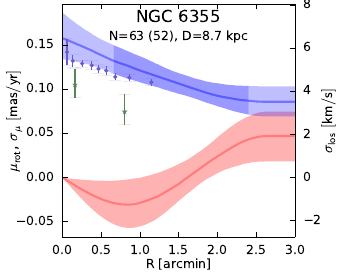}%
\includegraphics{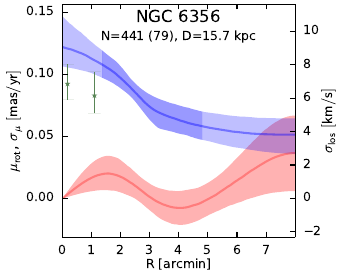}
\includegraphics{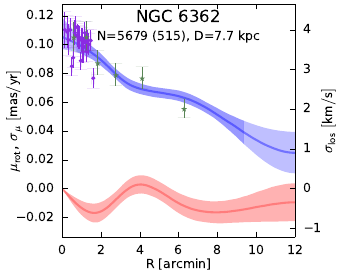}%
\includegraphics{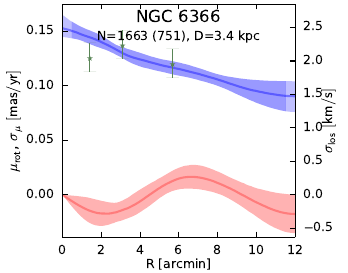}%
\includegraphics{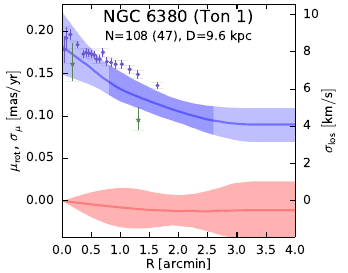}
\includegraphics{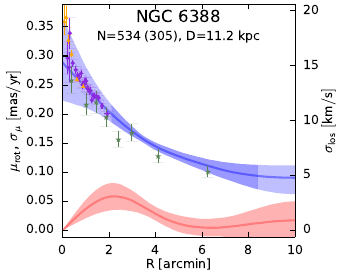}%
\includegraphics{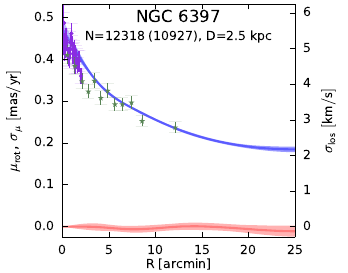}%
\includegraphics{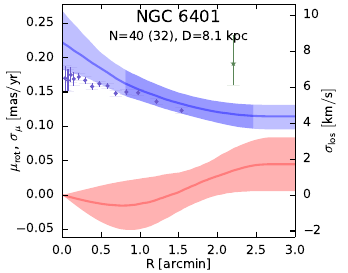}
\contcaption{}
\end{figure*}

\begin{figure*}
\includegraphics{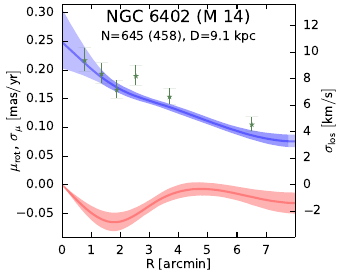}%
\includegraphics{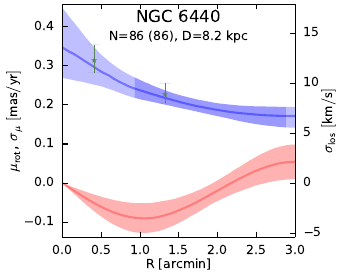}%
\includegraphics{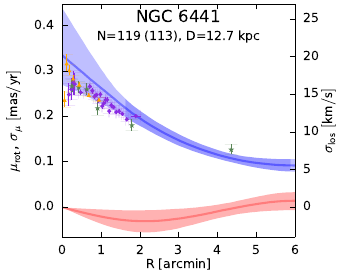}
\includegraphics{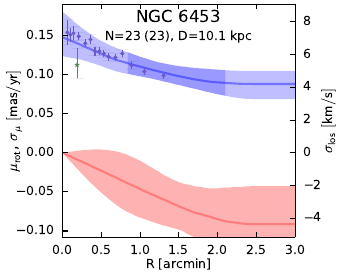}%
\includegraphics{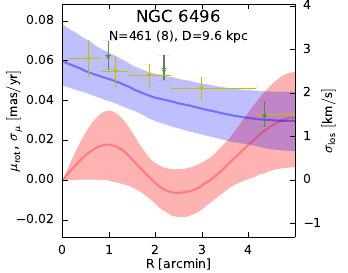}%
\includegraphics{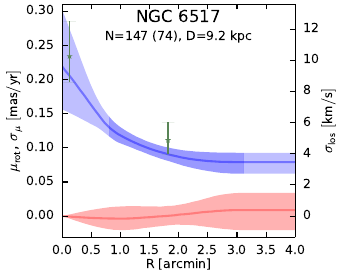}
\includegraphics{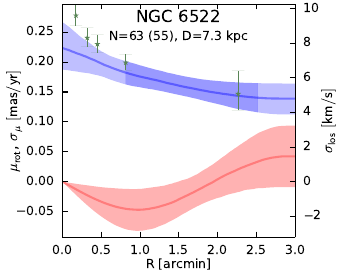}%
\includegraphics{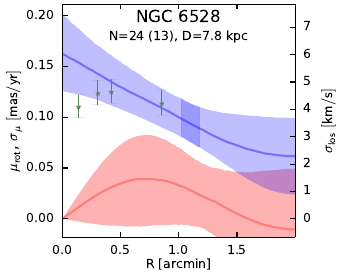}%
\includegraphics{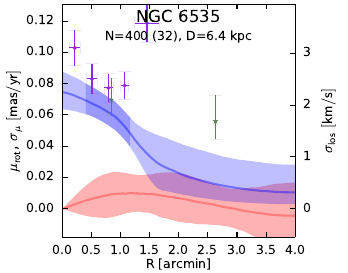}
\includegraphics{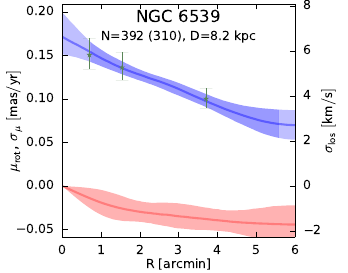}%
\includegraphics{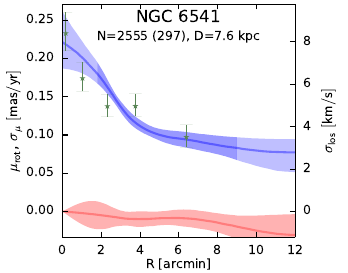}%
\includegraphics{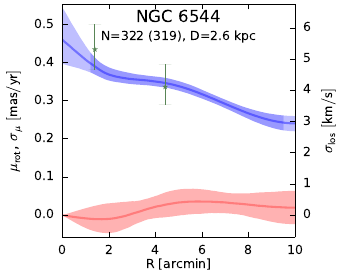}
\includegraphics{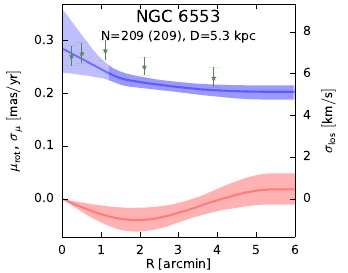}%
\includegraphics{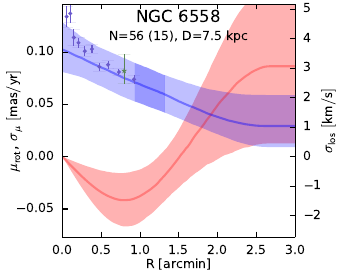}%
\includegraphics{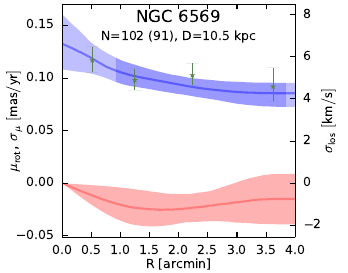}
\contcaption{}
\end{figure*}

\begin{figure*}
\includegraphics{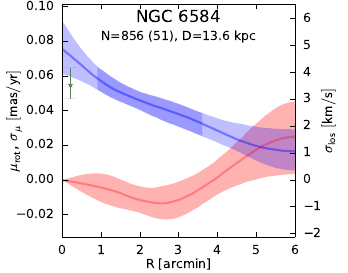}%
\includegraphics{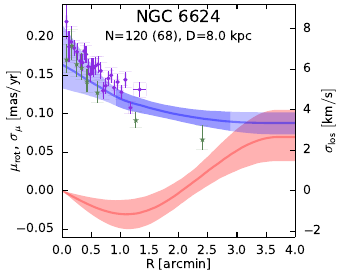}%
\includegraphics{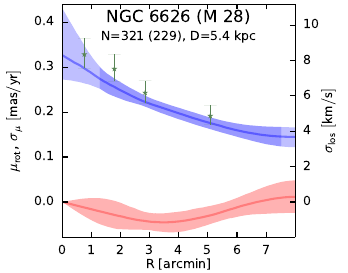}
\includegraphics{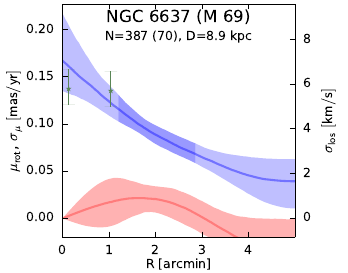}%
\includegraphics{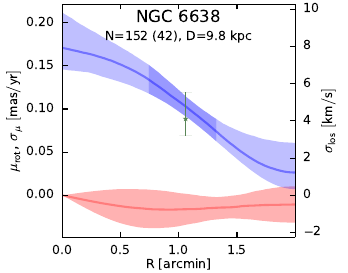}%
\includegraphics{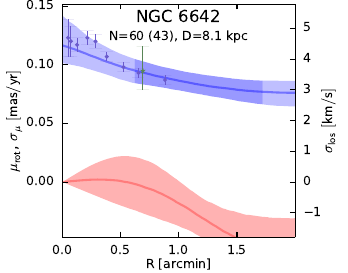}
\includegraphics{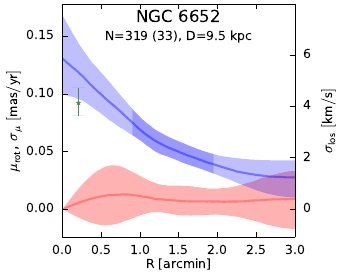}%
\includegraphics{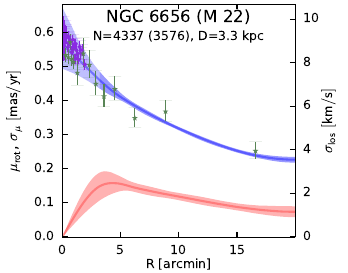}%
\includegraphics{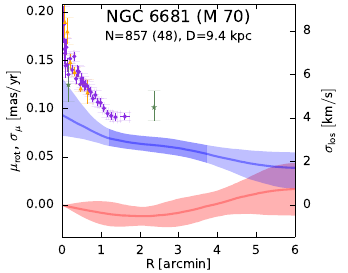}
\includegraphics{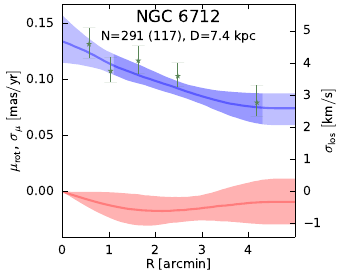}%
\includegraphics{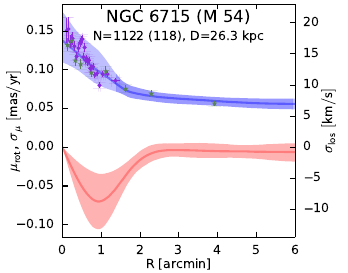}%
\includegraphics{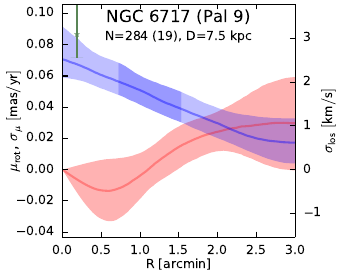}
\includegraphics{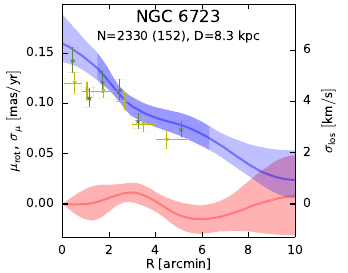}%
\includegraphics{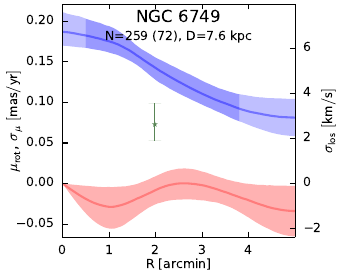}%
\includegraphics{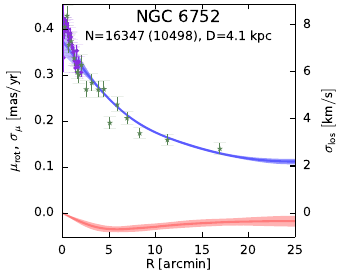}
\contcaption{}
\end{figure*}

\begin{figure*}
\includegraphics{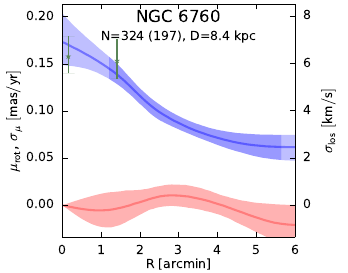}%
\includegraphics{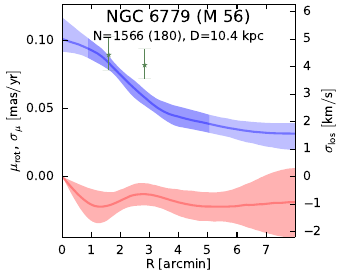}%
\includegraphics{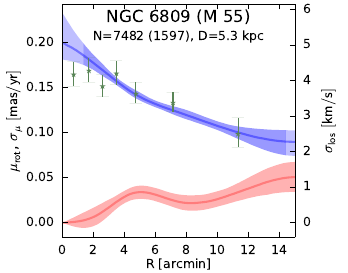}
\includegraphics{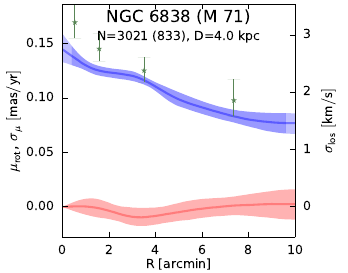}%
\includegraphics{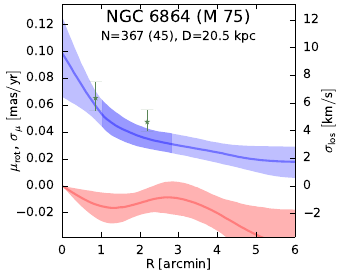}%
\includegraphics{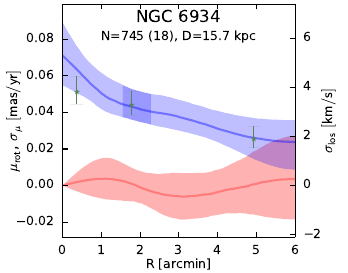}
\includegraphics{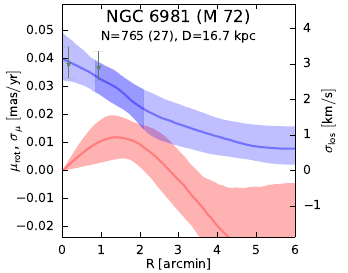}%
\includegraphics{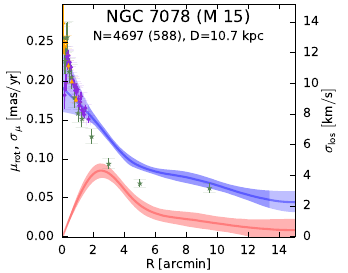}%
\includegraphics{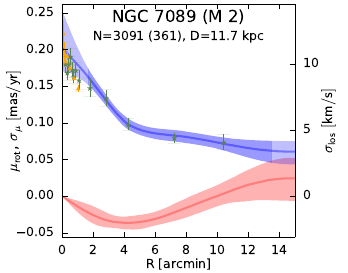}
\includegraphics{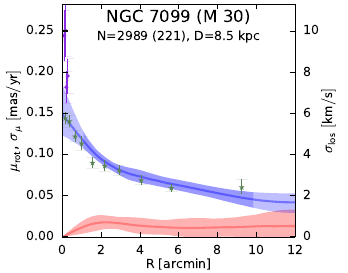}%
\includegraphics{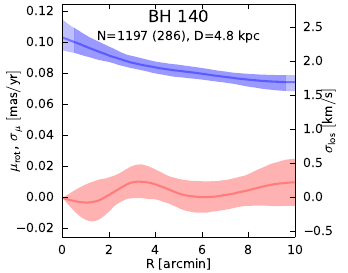}%
\includegraphics{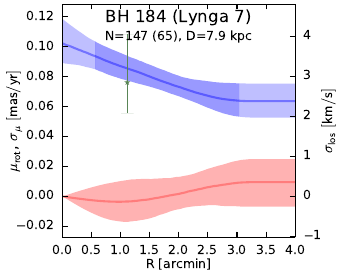}
\includegraphics{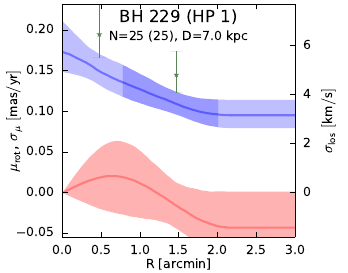}%
\includegraphics{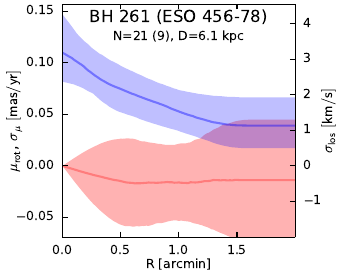}%
\includegraphics{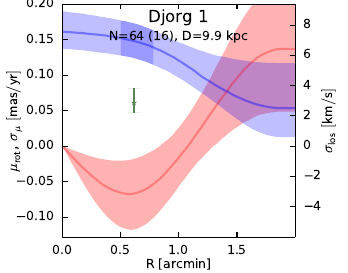}
\contcaption{}
\end{figure*}

\begin{figure*}
\includegraphics{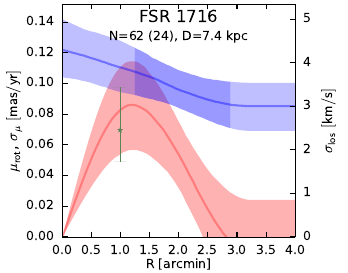}%
\includegraphics{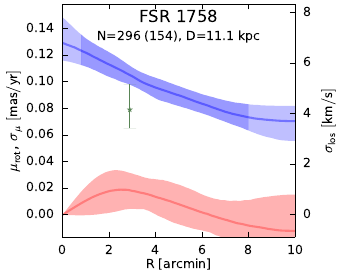}%
\includegraphics{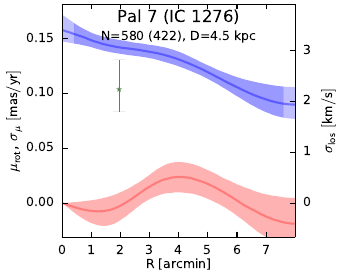}
\includegraphics{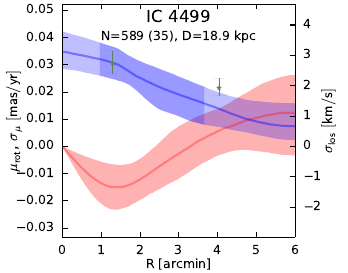}%
\includegraphics{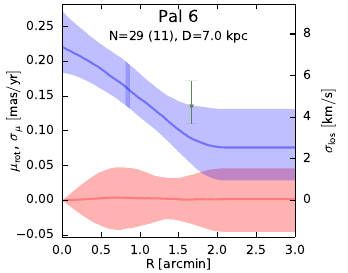}%
\includegraphics{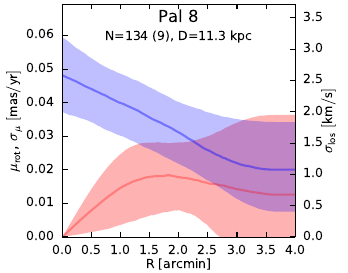}
\includegraphics{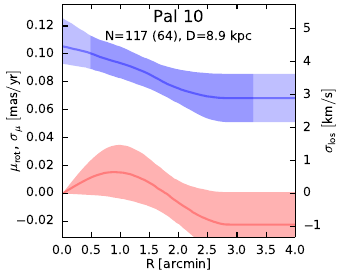}%
\includegraphics{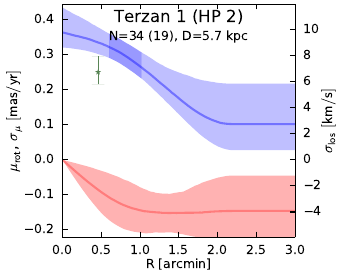}%
\includegraphics{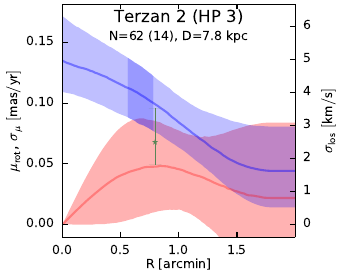}
\includegraphics{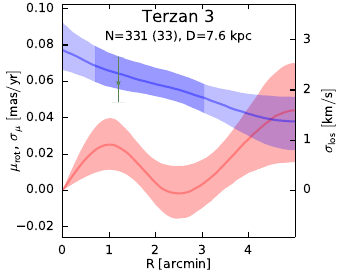}%
\includegraphics{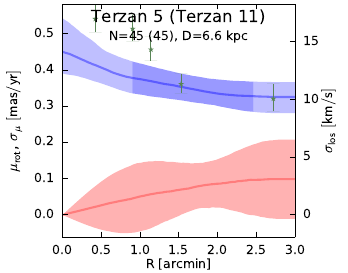}%
\includegraphics{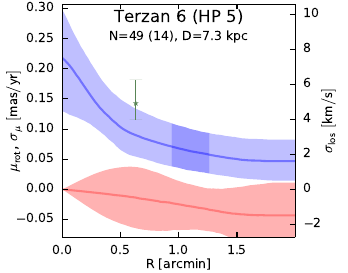}
\includegraphics{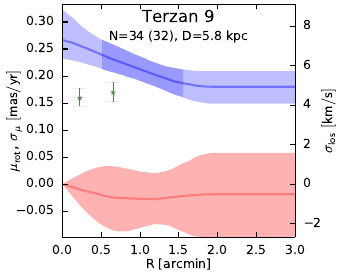}%
\includegraphics{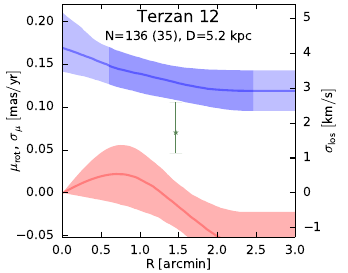}%
\includegraphics{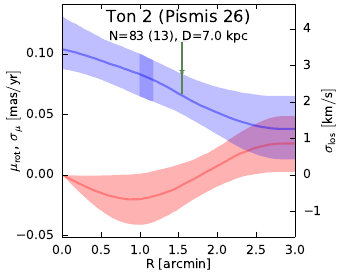}
\contcaption{}
\vspace*{10cm}
\end{figure*}

\begin{table*}
\caption{Catalogue of mean parallaxes and PM of Milky Way globular clusters,
derived from \Gaia EDR3 astrometry, taking into account spatially correlated systematic errors
(hence the uncertainty does not decrease below $0.01$~mas and $0.02$~\masyr).
Last two columns show the Plummer scale radius (in arcmin) and the number of member stars that pass all quality filters and are used to determine the cluster properties.
}  \label{tab:catalogue}
\begin{tabular}{p{3.0cm}rrrrrrrr}
\hline
Name & $\alpha$ [deg] & $\delta$ [deg] & $\overline{\mu_\alpha}$ [mas\,yr$^{-1}$] & $\overline{\mu_\delta}$ [mas\,yr$^{-1}$] & corr$_\mu$ & $\overline{\varpi}$ [mas] & $R_0$ [$'$] & \mbox{}\!\!$N_\mathrm{memb}$ \\
\hline
NGC 104 (47 Tuc)         & $  6.024$ & $-72.081$ & $  5.252\,\pm\,0.021$ & $ -2.551\,\pm\,0.021$ & $ 0.00$ & $ 0.232\,\pm\,0.009$ & $10.03$ & $39932$ \\
NGC 288                  & $ 13.188$ & $-26.583$ & $  4.164\,\pm\,0.024$ & $ -5.705\,\pm\,0.024$ & $ 0.01$ & $ 0.141\,\pm\,0.011$ & $2.79$ & $4689$ \\
NGC 362                  & $ 15.809$ & $-70.849$ & $  6.694\,\pm\,0.025$ & $ -2.535\,\pm\,0.024$ & $ 0.00$ & $ 0.114\,\pm\,0.011$ & $3.37$ & $3878$ \\
Whiting 1                & $ 30.737$ & $ -3.253$ & $ -0.228\,\pm\,0.065$ & $ -2.046\,\pm\,0.056$ & $ 0.03$ & $ 0.017\,\pm\,0.047$ & $0.52$ & $40$ \\
NGC 1261                 & $ 48.068$ & $-55.216$ & $  1.596\,\pm\,0.025$ & $ -2.064\,\pm\,0.025$ & $ 0.01$ & $ 0.068\,\pm\,0.011$ & $1.73$ & $1218$ \\
Pal 1                    & $ 53.333$ & $ 79.581$ & $ -0.252\,\pm\,0.034$ & $  0.007\,\pm\,0.037$ & $ 0.08$ & $ 0.112\,\pm\,0.023$ & $0.54$ & $92$ \\
E 1 (AM 1)               & $ 58.760$ & $-49.615$ & $  0.291\,\pm\,0.107$ & $ -0.177\,\pm\,0.086$ & $-0.22$ & $-0.015\,\pm\,0.062$ & $0.36$ & $58$ \\
Eridanus                 & $ 66.185$ & $-21.187$ & $  0.510\,\pm\,0.039$ & $ -0.301\,\pm\,0.041$ & $-0.09$ & $ 0.050\,\pm\,0.033$ & $0.52$ & $44$ \\
Pal 2                    & $ 71.525$ & $ 31.381$ & $  1.045\,\pm\,0.034$ & $ -1.522\,\pm\,0.031$ & $ 0.04$ & $ 0.042\,\pm\,0.021$ & $0.64$ & $180$ \\
NGC 1851                 & $ 78.528$ & $-40.047$ & $  2.145\,\pm\,0.024$ & $ -0.650\,\pm\,0.024$ & $-0.02$ & $ 0.088\,\pm\,0.011$ & $2.59$ & $2497$ \\
NGC 1904 (M 79)          & $ 81.044$ & $-24.524$ & $  2.469\,\pm\,0.025$ & $ -1.594\,\pm\,0.025$ & $ 0.00$ & $ 0.088\,\pm\,0.011$ & $1.90$ & $1556$ \\
NGC 2298                 & $102.248$ & $-36.005$ & $  3.320\,\pm\,0.025$ & $ -2.175\,\pm\,0.026$ & $ 0.01$ & $ 0.121\,\pm\,0.011$ & $1.39$ & $1047$ \\
NGC 2419                 & $114.535$ & $ 38.882$ & $  0.007\,\pm\,0.028$ & $ -0.523\,\pm\,0.026$ & $ 0.03$ & $ 0.003\,\pm\,0.017$ & $1.12$ & $333$ \\
Ko 2                     & $119.571$ & $ 26.255$ & $ -0.601\,\pm\,0.189$ & $ -0.025\,\pm\,0.129$ & $-0.24$ & $ 0.393\,\pm\,0.163$ & $1.57$ & $7$ \\
Pyxis                    & $136.991$ & $-37.221$ & $  1.030\,\pm\,0.032$ & $  0.138\,\pm\,0.035$ & $ 0.03$ & $ 0.016\,\pm\,0.022$ & $2.06$ & $69$ \\
NGC 2808                 & $138.013$ & $-64.863$ & $  0.994\,\pm\,0.024$ & $  0.273\,\pm\,0.024$ & $-0.01$ & $ 0.112\,\pm\,0.010$ & $4.08$ & $4740$ \\
E 3 (ESO 37-1)           & $140.238$ & $-77.282$ & $ -2.727\,\pm\,0.027$ & $  7.083\,\pm\,0.027$ & $ 0.00$ & $ 0.146\,\pm\,0.013$ & $1.70$ & $252$ \\
Pal 3                    & $151.383$ & $  0.072$ & $  0.086\,\pm\,0.060$ & $ -0.148\,\pm\,0.071$ & $-0.41$ & $-0.001\,\pm\,0.050$ & $0.71$ & $61$ \\
NGC 3201                 & $154.403$ & $-46.412$ & $  8.348\,\pm\,0.022$ & $ -1.958\,\pm\,0.022$ & $ 0.00$ & $ 0.222\,\pm\,0.010$ & $5.61$ & $11398$ \\
ESO 93-8                 & $169.925$ & $-65.220$ & $ -4.068\,\pm\,0.033$ & $  1.400\,\pm\,0.034$ & $-0.01$ & $ 0.101\,\pm\,0.018$ & $0.56$ & $32$ \\
Pal 4                    & $172.320$ & $ 28.974$ & $ -0.188\,\pm\,0.042$ & $ -0.476\,\pm\,0.041$ & $-0.14$ & $-0.025\,\pm\,0.033$ & $0.53$ & $96$ \\
Crater (Laevens 1)       & $174.067$ & $-10.877$ & $ -0.059\,\pm\,0.125$ & $ -0.116\,\pm\,0.116$ & $-0.23$ & $ 0.003\,\pm\,0.142$ & $0.51$ & $15$ \\
Bliss 1                  & $177.511$ & $-41.772$ & $ -2.340\,\pm\,0.042$ & $  0.138\,\pm\,0.038$ & $-0.16$ & $ 0.025\,\pm\,0.039$ & $0.62$ & $10$ \\
Ko 1                     & $179.827$ & $ 12.260$ & $ -1.513\,\pm\,0.135$ & $ -0.814\,\pm\,0.105$ & $-0.21$ & $-0.048\,\pm\,0.133$ & $0.30$ & $6$ \\
NGC 4147                 & $182.526$ & $ 18.543$ & $ -1.707\,\pm\,0.027$ & $ -2.090\,\pm\,0.027$ & $-0.02$ & $ 0.029\,\pm\,0.013$ & $0.89$ & $347$ \\
NGC 4372                 & $186.439$ & $-72.659$ & $ -6.409\,\pm\,0.024$ & $  3.297\,\pm\,0.024$ & $ 0.00$ & $ 0.187\,\pm\,0.010$ & $4.44$ & $2552$ \\
Rup 106                  & $189.667$ & $-51.150$ & $ -1.254\,\pm\,0.026$ & $  0.401\,\pm\,0.026$ & $ 0.03$ & $ 0.067\,\pm\,0.013$ & $1.23$ & $302$ \\
NGC 4590 (M 68)          & $189.867$ & $-26.744$ & $ -2.739\,\pm\,0.024$ & $  1.779\,\pm\,0.024$ & $ 0.00$ & $ 0.113\,\pm\,0.011$ & $2.61$ & $3284$ \\
BH 140                   & $193.473$ & $-67.177$ & $-14.848\,\pm\,0.024$ & $  1.224\,\pm\,0.024$ & $ 0.00$ & $ 0.219\,\pm\,0.011$ & $4.41$ & $1197$ \\
NGC 4833                 & $194.891$ & $-70.877$ & $ -8.377\,\pm\,0.025$ & $ -0.963\,\pm\,0.025$ & $-0.01$ & $ 0.164\,\pm\,0.011$ & $3.22$ & $2941$ \\
NGC 5024 (M 53)          & $198.230$ & $ 18.168$ & $ -0.133\,\pm\,0.024$ & $ -1.331\,\pm\,0.024$ & $ 0.00$ & $ 0.067\,\pm\,0.011$ & $3.02$ & $3398$ \\
NGC 5053                 & $199.113$ & $ 17.700$ & $ -0.329\,\pm\,0.025$ & $ -1.213\,\pm\,0.025$ & $-0.02$ & $ 0.050\,\pm\,0.011$ & $2.32$ & $1453$ \\
Kim 3                    & $200.688$ & $-30.601$ & $ -0.849\,\pm\,0.178$ & $  3.396\,\pm\,0.140$ & $-0.12$ & $ 0.103\,\pm\,0.159$ & $0.40$ & $9$ \\
NGC 5139 ($\omega$ Cen)  & $201.697$ & $-47.480$ & $ -3.250\,\pm\,0.022$ & $ -6.746\,\pm\,0.022$ & $ 0.01$ & $ 0.193\,\pm\,0.009$ & $14.62$ & $53127$ \\
NGC 5272 (M 3)           & $205.548$ & $ 28.377$ & $ -0.152\,\pm\,0.023$ & $ -2.670\,\pm\,0.022$ & $ 0.00$ & $ 0.110\,\pm\,0.010$ & $4.58$ & $8992$ \\
NGC 5286                 & $206.612$ & $-51.374$ & $  0.198\,\pm\,0.025$ & $ -0.153\,\pm\,0.025$ & $ 0.00$ & $ 0.088\,\pm\,0.011$ & $2.50$ & $847$ \\
AM 4                     & $209.090$ & $-27.167$ & $ -0.291\,\pm\,0.445$ & $ -2.512\,\pm\,0.344$ & $-0.36$ & $ 0.079\,\pm\,0.343$ & $0.39$ & $8$ \\
NGC 5466                 & $211.364$ & $ 28.534$ & $ -5.342\,\pm\,0.025$ & $ -0.822\,\pm\,0.024$ & $ 0.01$ & $ 0.057\,\pm\,0.011$ & $2.36$ & $2210$ \\
NGC 5634                 & $217.405$ & $ -5.976$ & $ -1.692\,\pm\,0.027$ & $ -1.478\,\pm\,0.026$ & $-0.01$ & $ 0.056\,\pm\,0.012$ & $0.97$ & $339$ \\
NGC 5694                 & $219.901$ & $-26.539$ & $ -0.464\,\pm\,0.029$ & $ -1.105\,\pm\,0.029$ & $-0.06$ & $ 0.034\,\pm\,0.017$ & $0.76$ & $149$ \\
IC 4499                  & $225.077$ & $-82.214$ & $  0.466\,\pm\,0.025$ & $ -0.489\,\pm\,0.025$ & $ 0.01$ & $ 0.054\,\pm\,0.011$ & $1.65$ & $589$ \\
Munoz 1                  & $225.450$ & $ 66.969$ & $ -0.100\,\pm\,0.203$ & $ -0.020\,\pm\,0.207$ & $-0.23$ & $ 0.221\,\pm\,0.136$ & $2.73$ & $5$ \\
NGC 5824                 & $225.994$ & $-33.068$ & $ -1.189\,\pm\,0.026$ & $ -2.234\,\pm\,0.026$ & $-0.02$ & $ 0.057\,\pm\,0.012$ & $1.51$ & $514$ \\
Pal 5                    & $229.019$ & $ -0.121$ & $ -2.730\,\pm\,0.028$ & $ -2.654\,\pm\,0.027$ & $ 0.00$ & $ 0.048\,\pm\,0.015$ & $3.17$ & $233$ \\
NGC 5897                 & $229.352$ & $-21.010$ & $ -5.422\,\pm\,0.025$ & $ -3.393\,\pm\,0.025$ & $-0.01$ & $ 0.105\,\pm\,0.011$ & $2.59$ & $2234$ \\
NGC 5904 (M 5)           & $229.638$ & $  2.081$ & $  4.086\,\pm\,0.023$ & $ -9.870\,\pm\,0.023$ & $-0.01$ & $ 0.141\,\pm\,0.010$ & $5.28$ & $9163$ \\
NGC 5927                 & $232.003$ & $-50.673$ & $ -5.056\,\pm\,0.025$ & $ -3.217\,\pm\,0.025$ & $-0.01$ & $ 0.127\,\pm\,0.011$ & $4.45$ & $1304$ \\
NGC 5946                 & $233.869$ & $-50.660$ & $ -5.331\,\pm\,0.028$ & $ -1.657\,\pm\,0.027$ & $-0.01$ & $ 0.112\,\pm\,0.012$ & $1.39$ & $180$ \\
BH 176                   & $234.781$ & $-50.050$ & $ -3.989\,\pm\,0.029$ & $ -3.057\,\pm\,0.029$ & $-0.02$ & $ 0.072\,\pm\,0.018$ & $1.34$ & $90$ \\
NGC 5986                 & $236.512$ & $-37.786$ & $ -4.192\,\pm\,0.026$ & $ -4.568\,\pm\,0.026$ & $-0.01$ & $ 0.111\,\pm\,0.011$ & $2.57$ & $1118$ \\
FSR 1716                 & $242.625$ & $-53.749$ & $ -4.354\,\pm\,0.033$ & $ -8.832\,\pm\,0.031$ & $ 0.00$ & $ 0.108\,\pm\,0.017$ & $1.96$ & $62$ \\
Pal 14 (Arp 1)           & $242.752$ & $ 14.958$ & $ -0.463\,\pm\,0.038$ & $ -0.413\,\pm\,0.038$ & $ 0.12$ & $ 0.035\,\pm\,0.032$ & $1.61$ & $80$ \\
BH 184 (Lynga 7)         & $242.765$ & $-55.318$ & $ -3.851\,\pm\,0.027$ & $ -7.050\,\pm\,0.027$ & $ 0.02$ & $ 0.121\,\pm\,0.012$ & $2.00$ & $147$ \\
NGC 6093 (M 80)          & $244.260$ & $-22.976$ & $ -2.934\,\pm\,0.027$ & $ -5.578\,\pm\,0.026$ & $-0.01$ & $ 0.102\,\pm\,0.011$ & $2.29$ & $1247$ \\
Ryu 059 (RLGC 1)         & $244.286$ & $-44.593$ & $  1.022\,\pm\,0.055$ & $  0.770\,\pm\,0.047$ & $ 0.01$ & $-0.066\,\pm\,0.046$ & $0.52$ & $77$ \\
NGC 6121 (M 4)           & $245.897$ & $-26.526$ & $-12.514\,\pm\,0.023$ & $-19.022\,\pm\,0.023$ & $-0.02$ & $ 0.556\,\pm\,0.010$ & $5.97$ & $5210$ \\
NGC 6101                 & $246.450$ & $-72.202$ & $  1.756\,\pm\,0.024$ & $ -0.258\,\pm\,0.025$ & $ 0.00$ & $ 0.084\,\pm\,0.011$ & $2.51$ & $2421$ \\
NGC 6144                 & $246.808$ & $-26.023$ & $ -1.744\,\pm\,0.026$ & $ -2.607\,\pm\,0.026$ & $ 0.00$ & $ 0.132\,\pm\,0.011$ & $1.78$ & $1586$ \\
NGC 6139                 & $246.918$ & $-38.849$ & $ -6.081\,\pm\,0.027$ & $ -2.711\,\pm\,0.026$ & $ 0.01$ & $ 0.112\,\pm\,0.011$ & $1.86$ & $336$ \\
Terzan 3                 & $247.167$ & $-35.353$ & $ -5.577\,\pm\,0.027$ & $ -1.760\,\pm\,0.026$ & $ 0.01$ & $ 0.133\,\pm\,0.012$ & $2.28$ & $331$ \\
NGC 6171 (M 107)         & $248.133$ & $-13.054$ & $ -1.939\,\pm\,0.025$ & $ -5.979\,\pm\,0.025$ & $ 0.00$ & $ 0.194\,\pm\,0.011$ & $2.39$ & $1668$ \\
\hline
\end{tabular}
\end{table*}
\begin{table*}
\begin{tabular}{p{3.0cm}rrrrrrrr}
\hline
Name & $\alpha$ [deg] & $\delta$ [deg] & $\overline{\mu_\alpha}$ [mas\,yr$^{-1}$] & $\overline{\mu_\delta}$ [mas\,yr$^{-1}$] & corr$_\mu$ & $\overline{\varpi}$ [mas] & $R_0$ [$'$] & \mbox{}\!\!$N_\mathrm{memb}$ \\
\hline
ESO 452-11 (1636-283)    & $249.854$ & $-28.399$ & $ -1.423\,\pm\,0.031$ & $ -6.472\,\pm\,0.030$ & $-0.06$ & $ 0.162\,\pm\,0.015$ & $0.62$ & $143$ \\
NGC 6205 (M 13)          & $250.422$ & $ 36.460$ & $ -3.149\,\pm\,0.023$ & $ -2.574\,\pm\,0.023$ & $-0.01$ & $ 0.127\,\pm\,0.010$ & $4.60$ & $9638$ \\
NGC 6229                 & $251.745$ & $ 47.528$ & $ -1.171\,\pm\,0.026$ & $ -0.467\,\pm\,0.027$ & $ 0.01$ & $ 0.041\,\pm\,0.012$ & $0.90$ & $256$ \\
NGC 6218 (M 12)          & $251.809$ & $ -1.949$ & $ -0.191\,\pm\,0.024$ & $ -6.802\,\pm\,0.024$ & $ 0.01$ & $ 0.208\,\pm\,0.011$ & $3.36$ & $6213$ \\
FSR 1735                 & $253.044$ & $-47.058$ & $ -4.439\,\pm\,0.054$ & $ -1.534\,\pm\,0.048$ & $-0.05$ & $-0.072\,\pm\,0.035$ & $1.10$ & $56$ \\
NGC 6235                 & $253.355$ & $-22.177$ & $ -3.931\,\pm\,0.027$ & $ -7.587\,\pm\,0.027$ & $-0.01$ & $ 0.089\,\pm\,0.012$ & $1.32$ & $270$ \\
NGC 6254 (M 10)          & $254.288$ & $ -4.100$ & $ -4.758\,\pm\,0.024$ & $ -6.597\,\pm\,0.024$ & $-0.02$ & $ 0.196\,\pm\,0.010$ & $4.14$ & $7269$ \\
NGC 6256                 & $254.886$ & $-37.121$ & $ -3.715\,\pm\,0.031$ & $ -1.637\,\pm\,0.030$ & $ 0.00$ & $ 0.162\,\pm\,0.013$ & $1.05$ & $102$ \\
Pal 15                   & $254.963$ & $ -0.539$ & $ -0.592\,\pm\,0.037$ & $ -0.901\,\pm\,0.034$ & $ 0.02$ & $ 0.024\,\pm\,0.025$ & $1.53$ & $125$ \\
NGC 6266 (M 62)          & $255.303$ & $-30.114$ & $ -4.978\,\pm\,0.026$ & $ -2.947\,\pm\,0.026$ & $ 0.01$ & $ 0.185\,\pm\,0.011$ & $5.87$ & $1100$ \\
NGC 6273 (M 19)          & $255.657$ & $-26.268$ & $ -3.249\,\pm\,0.026$ & $  1.660\,\pm\,0.025$ & $ 0.00$ & $ 0.142\,\pm\,0.011$ & $3.64$ & $1435$ \\
NGC 6284                 & $256.119$ & $-24.765$ & $ -3.200\,\pm\,0.029$ & $ -2.002\,\pm\,0.028$ & $ 0.00$ & $ 0.098\,\pm\,0.012$ & $1.29$ & $152$ \\
NGC 6287                 & $256.288$ & $-22.708$ & $ -5.010\,\pm\,0.029$ & $ -1.883\,\pm\,0.028$ & $ 0.01$ & $ 0.149\,\pm\,0.013$ & $1.72$ & $201$ \\
NGC 6293                 & $257.543$ & $-26.582$ & $  0.870\,\pm\,0.028$ & $ -4.326\,\pm\,0.028$ & $-0.01$ & $ 0.127\,\pm\,0.012$ & $1.80$ & $194$ \\
NGC 6304                 & $258.634$ & $-29.462$ & $ -4.070\,\pm\,0.029$ & $ -1.088\,\pm\,0.028$ & $ 0.01$ & $ 0.169\,\pm\,0.011$ & $2.11$ & $105$ \\
NGC 6316                 & $259.155$ & $-28.140$ & $ -4.969\,\pm\,0.031$ & $ -4.592\,\pm\,0.030$ & $ 0.01$ & $ 0.113\,\pm\,0.013$ & $1.51$ & $62$ \\
NGC 6341 (M 92)          & $259.281$ & $ 43.136$ & $ -4.935\,\pm\,0.024$ & $ -0.625\,\pm\,0.024$ & $ 0.00$ & $ 0.112\,\pm\,0.010$ & $3.29$ & $4365$ \\
NGC 6325                 & $259.497$ & $-23.766$ & $ -8.289\,\pm\,0.030$ & $ -9.000\,\pm\,0.029$ & $ 0.02$ & $ 0.162\,\pm\,0.013$ & $1.21$ & $106$ \\
NGC 6333 (M 9)           & $259.797$ & $-18.516$ & $ -2.180\,\pm\,0.026$ & $ -3.222\,\pm\,0.026$ & $ 0.02$ & $ 0.135\,\pm\,0.011$ & $3.24$ & $803$ \\
NGC 6342                 & $260.292$ & $-19.587$ & $ -2.903\,\pm\,0.027$ & $ -7.116\,\pm\,0.026$ & $ 0.00$ & $ 0.141\,\pm\,0.012$ & $1.54$ & $194$ \\
NGC 6356                 & $260.896$ & $-17.813$ & $ -3.750\,\pm\,0.026$ & $ -3.392\,\pm\,0.026$ & $ 0.00$ & $ 0.096\,\pm\,0.012$ & $2.32$ & $441$ \\
NGC 6355                 & $260.994$ & $-26.353$ & $ -4.738\,\pm\,0.031$ & $ -0.572\,\pm\,0.030$ & $ 0.02$ & $ 0.150\,\pm\,0.012$ & $1.43$ & $63$ \\
NGC 6352                 & $261.371$ & $-48.422$ & $ -2.158\,\pm\,0.025$ & $ -4.447\,\pm\,0.025$ & $-0.01$ & $ 0.190\,\pm\,0.011$ & $2.99$ & $2335$ \\
IC 1257                  & $261.785$ & $ -7.093$ & $ -1.007\,\pm\,0.040$ & $ -1.492\,\pm\,0.032$ & $ 0.10$ & $ 0.057\,\pm\,0.027$ & $0.36$ & $64$ \\
Terzan 2 (HP 3)          & $261.888$ & $-30.802$ & $ -2.170\,\pm\,0.041$ & $ -6.263\,\pm\,0.038$ & $ 0.05$ & $ 0.101\,\pm\,0.025$ & $0.65$ & $63$ \\
NGC 6366                 & $261.934$ & $ -5.080$ & $ -0.332\,\pm\,0.025$ & $ -5.160\,\pm\,0.024$ & $ 0.01$ & $ 0.285\,\pm\,0.011$ & $3.90$ & $1663$ \\
Terzan 4 (HP 4)          & $262.663$ & $-31.596$ & $ -5.462\,\pm\,0.060$ & $ -3.711\,\pm\,0.048$ & $ 0.19$ & $ 0.090\,\pm\,0.041$ & $0.99$ & $59$ \\
BH 229 (HP 1)            & $262.772$ & $-29.982$ & $  2.523\,\pm\,0.039$ & $-10.093\,\pm\,0.037$ & $ 0.04$ & $ 0.124\,\pm\,0.014$ & $0.93$ & $25$ \\
FSR 1758                 & $262.800$ & $-39.808$ & $ -2.881\,\pm\,0.026$ & $  2.519\,\pm\,0.025$ & $-0.01$ & $ 0.122\,\pm\,0.011$ & $4.05$ & $296$ \\
NGC 6362                 & $262.979$ & $-67.048$ & $ -5.506\,\pm\,0.024$ & $ -4.763\,\pm\,0.024$ & $ 0.00$ & $ 0.136\,\pm\,0.010$ & $3.02$ & $5678$ \\
Liller 1                 & $263.352$ & $-33.390$ & $ -5.403\,\pm\,0.109$ & $ -7.431\,\pm\,0.077$ & $ 0.30$ & $-0.089\,\pm\,0.083$ & $0.64$ & $83$ \\
NGC 6380 (Ton 1)         & $263.617$ & $-39.069$ & $ -2.183\,\pm\,0.031$ & $ -3.233\,\pm\,0.030$ & $ 0.01$ & $ 0.099\,\pm\,0.014$ & $0.96$ & $109$ \\
Terzan 1 (HP 2)          & $263.946$ & $-30.481$ & $ -2.806\,\pm\,0.055$ & $ -4.861\,\pm\,0.055$ & $ 0.01$ & $ 0.044\,\pm\,0.022$ & $0.92$ & $34$ \\
Ton 2 (Pismis 26)        & $264.044$ & $-38.553$ & $ -5.904\,\pm\,0.031$ & $ -0.755\,\pm\,0.029$ & $ 0.03$ & $ 0.143\,\pm\,0.015$ & $1.12$ & $83$ \\
NGC 6388                 & $264.072$ & $-44.736$ & $ -1.316\,\pm\,0.026$ & $ -2.709\,\pm\,0.026$ & $-0.01$ & $ 0.100\,\pm\,0.011$ & $3.34$ & $537$ \\
NGC 6402 (M 14)          & $264.400$ & $ -3.246$ & $ -3.590\,\pm\,0.025$ & $ -5.059\,\pm\,0.025$ & $ 0.01$ & $ 0.129\,\pm\,0.011$ & $3.00$ & $646$ \\
NGC 6401                 & $264.652$ & $-23.910$ & $ -2.748\,\pm\,0.035$ & $  1.444\,\pm\,0.034$ & $ 0.01$ & $ 0.149\,\pm\,0.013$ & $1.34$ & $40$ \\
NGC 6397                 & $265.175$ & $-53.674$ & $  3.260\,\pm\,0.023$ & $-17.664\,\pm\,0.022$ & $ 0.00$ & $ 0.416\,\pm\,0.010$ & $6.92$ & $12318$ \\
VVV CL002                & $265.276$ & $-28.845$ & $ -8.867\,\pm\,0.142$ & $  2.390\,\pm\,0.085$ & $ 0.17$ & $ 0.140\,\pm\,0.099$ & $0.41$ & $14$ \\
Pal 6                    & $265.926$ & $-26.223$ & $ -9.222\,\pm\,0.038$ & $ -5.347\,\pm\,0.036$ & $ 0.03$ & $ 0.104\,\pm\,0.017$ & $0.97$ & $29$ \\
NGC 6426                 & $266.228$ & $  3.170$ & $ -1.828\,\pm\,0.026$ & $ -2.999\,\pm\,0.026$ & $ 0.02$ & $ 0.051\,\pm\,0.012$ & $0.90$ & $130$ \\
Djorg 1                  & $266.868$ & $-33.066$ & $ -4.693\,\pm\,0.046$ & $ -8.468\,\pm\,0.041$ & $ 0.02$ & $ 0.094\,\pm\,0.027$ & $0.89$ & $64$ \\
Terzan 5 (Terzan 11)     & $267.020$ & $-24.779$ & $ -1.989\,\pm\,0.068$ & $ -5.243\,\pm\,0.066$ & $ 0.06$ & $ 0.145\,\pm\,0.022$ & $1.41$ & $45$ \\
NGC 6440                 & $267.220$ & $-20.360$ & $ -1.187\,\pm\,0.036$ & $ -4.020\,\pm\,0.035$ & $ 0.01$ & $ 0.149\,\pm\,0.013$ & $1.54$ & $86$ \\
NGC 6441                 & $267.554$ & $-37.051$ & $ -2.551\,\pm\,0.028$ & $ -5.348\,\pm\,0.028$ & $ 0.03$ & $ 0.083\,\pm\,0.011$ & $3.02$ & $119$ \\
Terzan 6 (HP 5)          & $267.693$ & $-31.275$ & $ -4.979\,\pm\,0.048$ & $ -7.431\,\pm\,0.039$ & $ 0.14$ & $ 0.076\,\pm\,0.033$ & $0.71$ & $49$ \\
NGC 6453                 & $267.715$ & $-34.599$ & $  0.203\,\pm\,0.036$ & $ -5.934\,\pm\,0.037$ & $-0.04$ & $ 0.098\,\pm\,0.013$ & $1.09$ & $23$ \\
UKS 1                    & $268.613$ & $-24.145$ & $ -2.023\,\pm\,0.107$ & $ -2.760\,\pm\,0.069$ & $-0.24$ & $ 0.056\,\pm\,0.091$ & $0.61$ & $112$ \\
VVV CL001                & $268.677$ & $-24.015$ & $ -3.486\,\pm\,0.155$ & $ -1.675\,\pm\,0.106$ & $ 0.08$ & $ 0.158\,\pm\,0.114$ & $0.39$ & $27$ \\
Gran 1                   & $269.653$ & $-32.020$ & $ -8.163\,\pm\,0.038$ & $ -8.045\,\pm\,0.036$ & $ 0.00$ & $ 0.076\,\pm\,0.025$ & $0.47$ & $19$ \\
Pfleiderer 2             & $269.664$ & $ -5.070$ & $ -2.784\,\pm\,0.034$ & $ -4.158\,\pm\,0.031$ & $ 0.09$ & $ 0.094\,\pm\,0.022$ & $1.41$ & $108$ \\
NGC 6496                 & $269.765$ & $-44.266$ & $ -3.060\,\pm\,0.027$ & $ -9.271\,\pm\,0.026$ & $ 0.01$ & $ 0.119\,\pm\,0.013$ & $1.73$ & $461$ \\
Terzan 9                 & $270.412$ & $-26.840$ & $ -2.121\,\pm\,0.052$ & $ -7.763\,\pm\,0.049$ & $ 0.03$ & $ 0.155\,\pm\,0.017$ & $0.80$ & $33$ \\
Djorg 2 (ESO 456-38)     & $270.455$ & $-27.826$ & $  0.662\,\pm\,0.042$ & $ -2.983\,\pm\,0.037$ & $ 0.00$ & $ 0.122\,\pm\,0.018$ & $0.73$ & $16$ \\
NGC 6517                 & $270.461$ & $ -8.959$ & $ -1.551\,\pm\,0.029$ & $ -4.470\,\pm\,0.028$ & $ 0.03$ & $ 0.115\,\pm\,0.012$ & $1.17$ & $147$ \\
Terzan 10                & $270.742$ & $-26.073$ & $ -6.827\,\pm\,0.059$ & $ -2.588\,\pm\,0.050$ & $ 0.01$ & $ 0.145\,\pm\,0.044$ & $1.06$ & $48$ \\
NGC 6522                 & $270.892$ & $-30.034$ & $  2.566\,\pm\,0.039$ & $ -6.438\,\pm\,0.036$ & $-0.02$ & $ 0.125\,\pm\,0.013$ & $1.09$ & $63$ \\
NGC 6535                 & $270.960$ & $ -0.298$ & $ -4.214\,\pm\,0.027$ & $ -2.939\,\pm\,0.026$ & $ 0.01$ & $ 0.163\,\pm\,0.012$ & $1.22$ & $400$ \\
NGC 6528                 & $271.207$ & $-30.056$ & $ -2.157\,\pm\,0.043$ & $ -5.649\,\pm\,0.039$ & $-0.03$ & $ 0.125\,\pm\,0.018$ & $0.70$ & $24$ \\
NGC 6539                 & $271.207$ & $ -7.586$ & $ -6.896\,\pm\,0.026$ & $ -3.537\,\pm\,0.026$ & $ 0.01$ & $ 0.131\,\pm\,0.011$ & $1.77$ & $392$ \\
NGC 6540 (Djorg 3)       & $271.536$ & $-27.765$ & $ -3.702\,\pm\,0.032$ & $ -2.791\,\pm\,0.032$ & $-0.02$ & $ 0.180\,\pm\,0.017$ & $0.65$ & $17$ \\
NGC 6544                 & $271.836$ & $-24.997$ & $ -2.304\,\pm\,0.031$ & $-18.604\,\pm\,0.030$ & $ 0.01$ & $ 0.399\,\pm\,0.011$ & $4.62$ & $322$ \\
NGC 6541                 & $272.010$ & $-43.715$ & $  0.287\,\pm\,0.025$ & $ -8.847\,\pm\,0.025$ & $ 0.00$ & $ 0.145\,\pm\,0.011$ & $3.63$ & $2555$ \\
ESO 280-06               & $272.275$ & $-46.423$ & $ -0.688\,\pm\,0.039$ & $ -2.777\,\pm\,0.033$ & $ 0.09$ & $ 0.041\,\pm\,0.026$ & $0.47$ & $45$ \\
NGC 6553                 & $272.323$ & $-25.909$ & $  0.344\,\pm\,0.030$ & $ -0.454\,\pm\,0.029$ & $ 0.01$ & $ 0.194\,\pm\,0.011$ & $3.47$ & $210$ \\
NGC 6558                 & $272.573$ & $-31.764$ & $ -1.720\,\pm\,0.036$ & $ -4.144\,\pm\,0.034$ & $ 0.04$ & $ 0.149\,\pm\,0.018$ & $0.94$ & $56$ \\
\hline
\end{tabular}
\end{table*}
\begin{table*}
\begin{tabular}{p{3.0cm}rrrrrrrr}
\hline
Name & $\alpha$ [deg] & $\delta$ [deg] & $\overline{\mu_\alpha}$ [mas\,yr$^{-1}$] & $\overline{\mu_\delta}$ [mas\,yr$^{-1}$] & corr$_\mu$ & $\overline{\varpi}$ [mas] & $R_0$ [$'$] & \mbox{}\!\!$N_\mathrm{memb}$ \\
\hline
Pal 7 (IC 1276)          & $272.684$ & $ -7.208$ & $ -2.553\,\pm\,0.026$ & $ -4.568\,\pm\,0.026$ & $ 0.01$ & $ 0.210\,\pm\,0.011$ & $3.31$ & $581$ \\
Terzan 12                & $273.066$ & $-22.742$ & $ -6.222\,\pm\,0.037$ & $ -3.052\,\pm\,0.034$ & $ 0.13$ & $ 0.187\,\pm\,0.020$ & $1.05$ & $136$ \\
NGC 6569                 & $273.412$ & $-31.827$ & $ -4.125\,\pm\,0.028$ & $ -7.354\,\pm\,0.028$ & $ 0.01$ & $ 0.112\,\pm\,0.012$ & $1.66$ & $102$ \\
BH 261 (ESO 456-78)      & $273.527$ & $-28.635$ & $  3.566\,\pm\,0.043$ & $ -3.590\,\pm\,0.037$ & $ 0.10$ & $ 0.191\,\pm\,0.023$ & $0.74$ & $21$ \\
NGC 6584                 & $274.657$ & $-52.216$ & $ -0.090\,\pm\,0.026$ & $ -7.202\,\pm\,0.025$ & $-0.01$ & $ 0.077\,\pm\,0.011$ & $1.76$ & $857$ \\
Mercer 5                 & $275.832$ & $-13.669$ & $ -3.965\,\pm\,0.114$ & $ -7.220\,\pm\,0.111$ & $ 0.31$ & $ 0.223\,\pm\,0.093$ & $2.02$ & $40$ \\
NGC 6624                 & $275.919$ & $-30.361$ & $  0.124\,\pm\,0.029$ & $ -6.936\,\pm\,0.029$ & $ 0.00$ & $ 0.130\,\pm\,0.012$ & $1.67$ & $120$ \\
NGC 6626 (M 28)          & $276.137$ & $-24.870$ & $ -0.278\,\pm\,0.028$ & $ -8.922\,\pm\,0.028$ & $-0.01$ & $ 0.200\,\pm\,0.011$ & $3.58$ & $321$ \\
NGC 6638                 & $277.734$ & $-25.497$ & $ -2.518\,\pm\,0.029$ & $ -4.076\,\pm\,0.029$ & $ 0.01$ & $ 0.115\,\pm\,0.012$ & $1.17$ & $152$ \\
NGC 6637 (M 69)          & $277.846$ & $-32.348$ & $ -5.034\,\pm\,0.028$ & $ -5.832\,\pm\,0.028$ & $ 0.04$ & $ 0.116\,\pm\,0.012$ & $1.97$ & $388$ \\
NGC 6642                 & $277.975$ & $-23.475$ & $ -0.173\,\pm\,0.030$ & $ -3.892\,\pm\,0.030$ & $ 0.00$ & $ 0.126\,\pm\,0.013$ & $1.05$ & $60$ \\
NGC 6652                 & $278.940$ & $-32.991$ & $ -5.484\,\pm\,0.027$ & $ -4.274\,\pm\,0.027$ & $ 0.01$ & $ 0.119\,\pm\,0.012$ & $1.33$ & $319$ \\
NGC 6656 (M 22)          & $279.100$ & $-23.905$ & $  9.851\,\pm\,0.023$ & $ -5.617\,\pm\,0.023$ & $ 0.00$ & $ 0.306\,\pm\,0.010$ & $8.08$ & $4337$ \\
Pal 8                    & $280.375$ & $-19.826$ & $ -1.987\,\pm\,0.027$ & $ -5.694\,\pm\,0.027$ & $ 0.01$ & $ 0.115\,\pm\,0.012$ & $1.16$ & $133$ \\
NGC 6681 (M 70)          & $280.803$ & $-32.292$ & $  1.431\,\pm\,0.027$ & $ -4.744\,\pm\,0.026$ & $ 0.01$ & $ 0.121\,\pm\,0.012$ & $2.01$ & $857$ \\
Ryu 879 (RLGC 2)         & $281.367$ & $ -5.192$ & $ -2.396\,\pm\,0.077$ & $ -1.794\,\pm\,0.069$ & $ 0.39$ & $ 0.084\,\pm\,0.074$ & $0.32$ & $34$ \\
NGC 6712                 & $283.268$ & $ -8.706$ & $  3.363\,\pm\,0.027$ & $ -4.436\,\pm\,0.027$ & $ 0.02$ & $ 0.146\,\pm\,0.011$ & $2.00$ & $291$ \\
NGC 6715 (M 54)          & $283.764$ & $-30.480$ & $ -2.679\,\pm\,0.025$ & $ -1.387\,\pm\,0.025$ & $ 0.00$ & $ 0.053\,\pm\,0.011$ & $3.36$ & $1122$ \\
NGC 6717 (Pal 9)         & $283.775$ & $-22.701$ & $ -3.125\,\pm\,0.027$ & $ -5.008\,\pm\,0.027$ & $ 0.03$ & $ 0.117\,\pm\,0.012$ & $1.21$ & $284$ \\
NGC 6723                 & $284.888$ & $-36.632$ & $  1.028\,\pm\,0.025$ & $ -2.418\,\pm\,0.025$ & $ 0.00$ & $ 0.132\,\pm\,0.011$ & $2.93$ & $2330$ \\
NGC 6749                 & $286.314$ & $  1.901$ & $ -2.829\,\pm\,0.028$ & $ -6.006\,\pm\,0.027$ & $ 0.01$ & $ 0.138\,\pm\,0.012$ & $2.15$ & $259$ \\
NGC 6752                 & $287.717$ & $-59.985$ & $ -3.161\,\pm\,0.022$ & $ -4.027\,\pm\,0.022$ & $-0.01$ & $ 0.254\,\pm\,0.010$ & $6.08$ & $16348$ \\
NGC 6760                 & $287.800$ & $  1.030$ & $ -1.107\,\pm\,0.026$ & $ -3.615\,\pm\,0.026$ & $ 0.02$ & $ 0.132\,\pm\,0.011$ & $2.30$ & $324$ \\
NGC 6779 (M 56)          & $289.148$ & $ 30.183$ & $ -2.018\,\pm\,0.025$ & $  1.618\,\pm\,0.025$ & $ 0.01$ & $ 0.091\,\pm\,0.011$ & $2.03$ & $1567$ \\
Terzan 7                 & $289.433$ & $-34.658$ & $ -3.002\,\pm\,0.029$ & $ -1.651\,\pm\,0.029$ & $ 0.05$ & $ 0.050\,\pm\,0.015$ & $0.83$ & $116$ \\
Pal 10                   & $289.509$ & $ 18.572$ & $ -4.322\,\pm\,0.029$ & $ -7.173\,\pm\,0.029$ & $ 0.02$ & $ 0.112\,\pm\,0.014$ & $1.49$ & $117$ \\
Arp 2                    & $292.184$ & $-30.356$ & $ -2.331\,\pm\,0.031$ & $ -1.475\,\pm\,0.029$ & $ 0.05$ & $ 0.026\,\pm\,0.021$ & $1.66$ & $156$ \\
NGC 6809 (M 55)          & $294.999$ & $-30.965$ & $ -3.432\,\pm\,0.024$ & $ -9.311\,\pm\,0.024$ & $ 0.01$ & $ 0.209\,\pm\,0.010$ & $4.70$ & $7483$ \\
Terzan 8                 & $295.435$ & $-33.999$ & $ -2.496\,\pm\,0.027$ & $ -1.581\,\pm\,0.026$ & $-0.02$ & $ 0.050\,\pm\,0.013$ & $1.99$ & $354$ \\
Pal 11                   & $296.310$ & $ -8.007$ & $ -1.766\,\pm\,0.030$ & $ -4.971\,\pm\,0.028$ & $ 0.04$ & $ 0.109\,\pm\,0.017$ & $1.18$ & $172$ \\
NGC 6838 (M 71)          & $298.444$ & $ 18.779$ & $ -3.416\,\pm\,0.025$ & $ -2.656\,\pm\,0.024$ & $ 0.00$ & $ 0.251\,\pm\,0.011$ & $3.22$ & $3021$ \\
NGC 6864 (M 75)          & $301.520$ & $-21.921$ & $ -0.598\,\pm\,0.026$ & $ -2.810\,\pm\,0.026$ & $ 0.02$ & $ 0.069\,\pm\,0.012$ & $1.31$ & $367$ \\
NGC 6934                 & $308.547$ & $  7.404$ & $ -2.655\,\pm\,0.026$ & $ -4.689\,\pm\,0.026$ & $ 0.01$ & $ 0.078\,\pm\,0.012$ & $1.58$ & $745$ \\
NGC 6981 (M 72)          & $313.365$ & $-12.537$ & $ -1.274\,\pm\,0.026$ & $ -3.361\,\pm\,0.026$ & $ 0.00$ & $ 0.084\,\pm\,0.012$ & $1.34$ & $765$ \\
NGC 7006                 & $315.372$ & $ 16.187$ & $ -0.128\,\pm\,0.027$ & $ -0.633\,\pm\,0.027$ & $ 0.00$ & $ 0.035\,\pm\,0.015$ & $0.61$ & $174$ \\
Laevens 3                & $316.729$ & $ 14.984$ & $  0.172\,\pm\,0.101$ & $ -0.666\,\pm\,0.080$ & $-0.02$ & $ 0.126\,\pm\,0.089$ & $0.52$ & $21$ \\
Segue 3                  & $320.379$ & $ 19.117$ & $ -0.981\,\pm\,0.121$ & $ -1.667\,\pm\,0.081$ & $-0.31$ & $-0.005\,\pm\,0.117$ & $0.49$ & $21$ \\
NGC 7078 (M 15)          & $322.493$ & $ 12.167$ & $ -0.659\,\pm\,0.024$ & $ -3.803\,\pm\,0.024$ & $ 0.01$ & $ 0.097\,\pm\,0.010$ & $4.10$ & $4699$ \\
NGC 7089 (M 2)           & $323.363$ & $ -0.823$ & $  3.435\,\pm\,0.025$ & $ -2.159\,\pm\,0.024$ & $ 0.01$ & $ 0.082\,\pm\,0.011$ & $3.52$ & $3091$ \\
NGC 7099 (M 30)          & $325.092$ & $-23.180$ & $ -0.737\,\pm\,0.025$ & $ -7.299\,\pm\,0.024$ & $ 0.01$ & $ 0.136\,\pm\,0.011$ & $2.77$ & $2990$ \\
Pal 12                   & $326.662$ & $-21.253$ & $ -3.220\,\pm\,0.029$ & $ -3.333\,\pm\,0.028$ & $ 0.08$ & $ 0.050\,\pm\,0.018$ & $0.95$ & $185$ \\
Pal 13                   & $346.685$ & $ 12.772$ & $  1.748\,\pm\,0.049$ & $  0.104\,\pm\,0.047$ & $-0.02$ & $-0.037\,\pm\,0.034$ & $0.66$ & $72$ \\
NGC 7492                 & $347.111$ & $-15.611$ & $  0.756\,\pm\,0.028$ & $ -2.320\,\pm\,0.028$ & $ 0.03$ & $ 0.073\,\pm\,0.014$ & $1.13$ & $190$ \\
\hline
\end{tabular}
\end{table*}

\end{document}